\documentclass[fleqn,usenatbib]{mnras}

\usepackage{newtxtext,newtxmath}

\usepackage[T1]{fontenc}

\DeclareRobustCommand{\VAN}[3]{#2}
\let\VANthebibliography\thebibliography
\def\thebibliography{\DeclareRobustCommand{\VAN}[3]{##3}\VANthebibliography}

\usepackage{graphicx}	
\usepackage{amsmath}	

\title[Fitting Eclipse Maps of Exoplanets]{Identifying and Fitting Eclipse Maps of Exoplanets with Cross-Validation}

\author[M. Hammond et al.]{
Mark Hammond,$^{1}$\thanks{E-mail: mark.hammond@physics.ox.ac.uk}
Neil T. Lewis,$^{2}$
Sasha Boone,$^{1}$
Xueqing Chen,$^{3}$
João M. Mendonça,$^{4}$
\newauthor
Vivien Parmentier,$^{6}$
Jake Taylor,$^{5}$
Taylor Bell,$^{7,8}$
Leonardo dos Santos,$^{9}$
Nicolas Crouzet,$^{10}$
\newauthor
Laura Kreidberg,$^{11}$
Michael Radica,$^{12}$
and Michael Zhang$^{13}$
\\
$^{1}$Atmospheric, Oceanic and Planetary Physics, Department of Physics, University of Oxford, Oxford, UK\\
$^{2}$Department of Physics and Astronomy, University of Exeter, Exeter, UK\\
$^{3}$Institute for Astronomy, University of Edinburgh, Royal Observatory, Blackford Hill, Edinburgh, EH9 3HJ, UK\\
$^{4}$National Space Institute, Technical University of Denmark, Elektrovej, 2800 Kgs. Lyngby, Denmark\\
$^{5}$Astrophysics, Department of Physics, University of Oxford, Oxford, UK\\
$^{6}$Université Côte d’Azur, Av. Valrose, 06000 Nice, France\\
$^{7}$BAER Institute, NASA Ames Research Center, Moffett Field, CA, USA\\
$^{8}$Space Science and Astrobiology Division, NASA Ames Research Center, Moffett Field, CA, USA\\
$^{9}$Space Telescope Science Institute, Baltimore, MD, USA\\
$^{10}$Leiden Observatory, Leiden University, P.O. Box 9513, 2300 RA Leiden, The Netherlands\\
$^{11}$Max Planck Institute for Astronomy, Heidelberg, Germany\\
$^{12}$Trottier Institute for Research on Exoplanets, Université de Montréal, Montréal, Québec, Canada\\
$^{13}$Department of Astronomy and Astrophysics, The University of Chicago, Chicago, IL 60637, USA\\
}

\date{Accepted XXX. Received YYY; in original form ZZZ}

\pubyear{\the\year{}}

\begin{document}
\label{firstpage}
\pagerange{\pageref{firstpage}--\pageref{lastpage}}
\maketitle

\begin{abstract}
Eclipse mapping uses the shape of the eclipse of an exoplanet to measure its two-dimensional structure. Light curves are mostly composed of longitudinal information, with the latitudinal information only contained in the brief ingress and egress of the eclipse. This imbalance can lead to a spuriously confident map, where the longitudinal structure is constrained by out-of-eclipse data and the latitudinal structure is wrongly determined by the priors on the map. We present a new method to address this issue. The method tests for the presence of an eclipse mapping signal by using k-fold cross-validation to compare the performance of a simple mapping model to the null hypothesis of a uniform disk. If a signal is found, the method fits a map with more degrees of freedom, optimising its information content. The information content is varied by penalising the model likelihood by a factor proportional to the spatial entropy of the map, optimised by cross-validation. We demonstrate this method for simulated datasets then apply it to three observational datasets. The method identifies an eclipse mapping signal for JWST MIRI/LRS observations of WASP-43b but does not identify a signal for JWST NIRISS/SOSS observations of WASP-18b or Spitzer Space Telescope observations of HD 189733b. It is possible to fit eclipse maps to these datasets, but we suggest that these maps are overfitting the eclipse shape. We fit a new map with more spatial freedom to the WASP-43b dataset and show a flatter east-west structure than previously derived.
\end{abstract}




\section{Introduction} 

Planetary atmospheres are controlled by multi-dimensional processes. Observations of transiting exoplanets are generally limited to measurements of point sources, where the flux from the entire stellar and planetary system is observed at once. For such a system, eclipse mapping is currently the only method by which both two-dimensional longitudinal (east-west) and latitudinal (north-south) information can be measured over the surface of an exoplanet. This information is derived from the shape of the eclipse of an exoplanet by its star. The eclipse shape is a function of the spatial distribution of flux from the planet and of the geometry of the coverage of the planet by its star. If we know the geometry of the eclipse coverage, or can fit a model of it, we can derive a map of the flux from the planet.

Eclipse mapping has been used in other contexts to derive images of the surface brightness of accretion disks \citep{horne1985images} and maps of the single-scattering albedo of Pluto and Charon \citep{buie1992albedo}. \citet{rauscher2007toward} showed the possibility of mapping exoplanets with sufficiently high-precision observations, highlighting the suitability of JWST in particular. Eclipse mapping has previously been applied to observations of three ``hot Jupiter'' exoplanets: HD 189733b with the Spitzer Space Telescope \citep{de2012towards,majeau2012two}, WASP-18b with JWST NIRISS/SOSS \citep{coulombe2023broadband}, and WASP-43b with JWST MIRI/LRS \citep{hammond2024wasp43b}. \citet{majeau2012two} used spherical harmonics up to first order to map HD 189733b. They found that the brightest part of the planet was shifted slightly eastwards, and that its latitudinal position was consistent with the equator. \citet{de2012towards} used more flexible mapping functions to derive a more localised hot-spot from this same dataset and highlighted degeneracies between this map and the orbital parameters of the planet. \citep{coulombe2023broadband} derived a map of WASP-18b with a notably uniform longitudinal structure near the substellar point, but they could not find evidence for latitudinal information in the dataset. \citet{hammond2024wasp43b} detected latitudinal information in the eclipse shape of WASP-43b and found an eastward longitudinal hot-spot shift.

Deriving reliable eclipse maps from observational datasets has two intrinsic issues. First, the problem is ill-posed -- one spatial pattern can produce the same flux time-series as another, or even no flux at all \citep{cowan2008inverting,challener2023eclipse}. Alongside the uncertainty inherent in observational data, this means that an observed light curve implies a range of possible maps instead of a single unique solution. Second, the data that contain two-dimensional information are normally a very small part of an overall time-series observation -- the ingress and egress of an eclipse being typically at least an order of magnitude shorter than the whole orbit of a planet. These issues lead to two statistical challenges.

The first challenge is to identify when it is justifiable to fit an eclipse map to a particular observation. In general, an eclipse map model has more degrees of freedom than a model of a planet with uniform emission. It will, therefore, normally produce a better fit to the data in an eclipse, even when the deviations from the shape of a uniform eclipse are driven by random noise, or when unresolved systematic errors produce an eclipse shape that does not depend on the true emission map. This means that it is easy to fit spurious eclipse maps to noisy or inaccurate datasets. Section \ref{sec:methods} therefore presents a statistical test using k-fold cross-validation to determine when an ``eclipse mapping signal'' is present. This statistical test avoids both underfitting (where the model cannot match the observations), and overfitting (where the model derives spurious information from noise). \citet{welbanks2023application} shows a similar method for fitting spectroscopic models to exoplanet observations using leave-one-out cross validation. \citet{challener2023bringing} showed the application of leave-one-out cross-validation to eclipse mapping, but we chose not to use the leave-one-out method as we seek to test the predictive power of a model for a longer duration of omitted data.

The second statistical challenge is how much complexity to allow in the fitted map, if an eclipse mapping signal is indeed present. Fitting with a high degree of spatial freedom results in highly uncertain maps with spurious small-scale features due to uncertainty and degeneracy in the observational data. \citet{de2012towards} investigated this issue issue, showing how fitting an eclipse map with a model with more small-scale freedom (a single hot-spot of variable size) resulted in greater uncertainty than a model with only large-scale freedom (a low-order spherical harmonic map). \citet{rauscher2018more} developed the method of ``eigenmapping'' to address this issue, orthogonalising the light-curves used to fit an observation, and then determining the number of corresponding maps used to fit the data by optimising the Bayesian Information Criterion. In Section \ref{sec:methods} we present a new eclipse mapping method to derive the appropriate spatial complexity of the fitted map, again using k-fold cross-validation. This method is based on the pixel sampling and entropy maximisation methods presented in \citet{horne1985images} and \citet{chen2024map} for mapping of accretion disks and brown dwarfs.




We demonstrate these methods with simulated data in Section \ref{sec:results_sim}, showing that they correctly identify when fitting a simple low-order eclipse map is justified. We show that the method can then identify the appropriate information content for a map composed of higher-order spherical harmonics, depending on the spatial scale of the true map and the precision of the data. We then demonstrate this method for real observational data in Section \ref{sec:results_obs}, applying it to observations of WASP-43b with JWST MIRI/LRS, WASP-18b with JWST NIRISS/SOSS, and HD 189733b with the Spitzer Space Telescope. The method identifies a clear eclipse mapping signal for observations of WASP-43b with JWST MIRI/LRS, consistent with the findings of \citet{hammond2024wasp43b}, and fits a new map optimised by our new method. Our method does not identify a significant eclipse mapping signal for observations of WASP-18b with NIRISS SOSS or the Spitzer Space Telescope observations of HD 189733b; this differs from the conclusions of \citet{coulombe2023broadband}, \citet{majeau2012two}, and \citet{de2012towards}.




We conclude that our proposed method can reliably determine the presence of an eclipse mapping signal in observational data, and then determine the appropriate spatial scale of the map to fit to this data, addressing the two statistical issues presented above.

\begin{figure*}
\centering 
  \includegraphics[width=\textwidth]{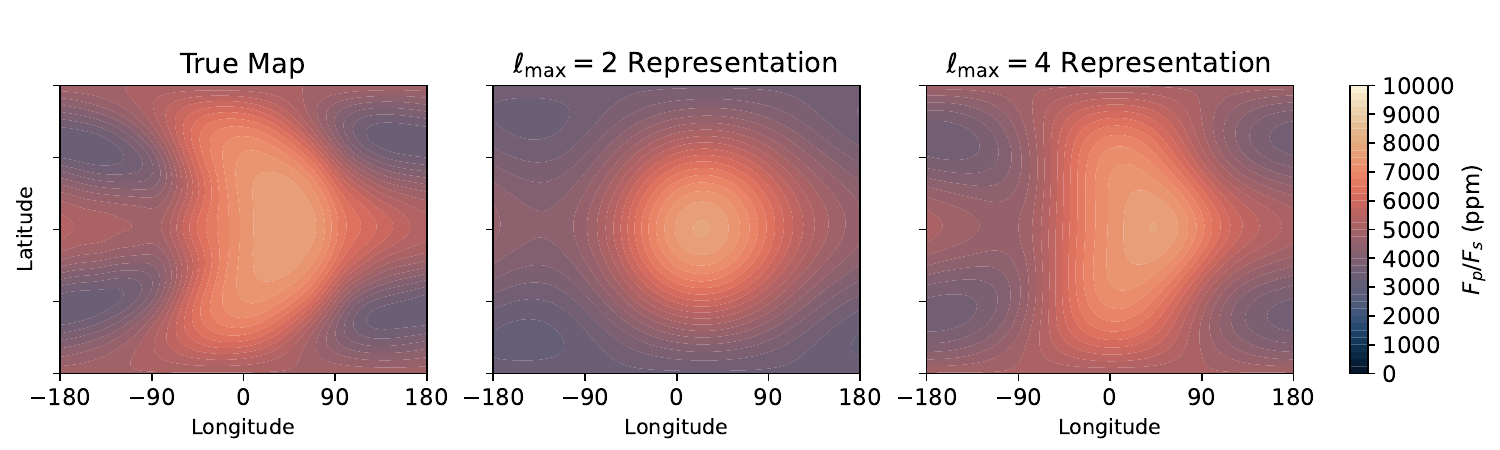}%
\caption{First panel: the emission map from the \texttt{THOR} GCM used to simulate the time-series observations in Figures \ref{fig:sim_w43b_good} and \ref{fig:sim_w43b_bad}. Second panel: the true map represented by $\ell_{\mathrm{max}}=2$ spherical harmonics, showing how these are not sufficient to capture its spatial structure. Third panel: the true map represented by $\ell_{\mathrm{max}}=4$ spherical harmonics, showing how this captures most of the true structure. The brightest point of the true map is at $40^{\circ}$ east of the substellar point. The brightest point of the $\ell_{\mathrm{max}}=2$ representation is incorrect, at $24^{\circ}$ east. The $\ell_{\mathrm{max}}=4$ representation is more accurate, with its brightest point at $43^{\circ}$ east.}\label{fig:gcm_true_l2_l4}
\end{figure*}

\section{Methods}\label{sec:methods}

This section describes the methods we use to simulate and retrieve eclipse mapping data. It lists the models used to simulate observations of thermal emission light curves, our sources of real observational light curves, the models used to fit these light curves with eclipse maps, and the cross-validation metric we propose to compare the quality of different models.

\subsection{Simulated time-series thermal emission}\label{sec:methods:simulation}

We simulate time-series thermal emission to represent the typical precision and cadence of observations suitable for eclipse mapping with JWST. We set up our simulation to emulate observations of the hot Jupiter WASP-43b \citep{hellier2011wasp43b} with the MIRI/LRS instrument on JWST \citep{bell2024nightside}. Our conclusions do not depend on this particular choice, and we vary our model choices later. WASP-43b orbits a K7, 0.717 M\textsubscript{\(\odot\)} star with an orbital period of 19.2 hours, a mass of 2.034 $M_{J}$, and a semi-major axis of 0.0153 AU \citep{gillon2012wasp43b}. It has an estimated equilibrium temperature of 1440 K \citep{blecic2014wasp43b}.

We use a simulation from the General Circulation Model (GCM) \texttt{THOR} \citep{mendoncca2016thor, deitrick2020thor} of a cloudless atmosphere on WASP-43b to model observations.  This simulation was previously used to study the atmospheric temperature structure, cloud cover, and chemistry of WASP-43b \citep{mendoncca2018revisiting, mendoncca2018chemistry}. \texttt{THOR} is based on a dynamical core that solves the non-hydrostatic compressible Euler equations on an icosahedral grid. The simulation uses a simple two-band formulation of radiative transfer calibrated to reproduce the results from more complex non-grey models of WASP-43b \citep{kataria2015wasp43b, malik2017helios}. More details about the model configuration can be found in \citet{mendoncca2018revisiting}. We are not concerned about the time-varying behaviour of the atmosphere (which is small in any case) as we are using this as an exemplar case, so we post-process a single snapshot of the GCM temperature field.

To simulate the observed thermal emission in the MIRI/LRS bandpass, the GCM simulation was post-processed using the HELIOS code \citep{malik2017helios}  to produce a map of the planetary thermal flux averaged over the entire bandpass, normalised by a stellar spectrum from the PHOENIX model database \citep{allard1995model,husser2013new} as described in \citet{mendoncca2018chemistry}. We do not represent limb darkening for this latitude-longitude map, as we fit a two-dimensional map that varies only due to the rotation and occultation of the planet.

This simulation results in the map of thermal emission shown in the left-hand panel of Figure \ref{fig:gcm_true_l2_l4}. The figure also shows representations of this ``true map'' using spherical harmonics up to a given order $\ell_{\mathrm{max}}$, calculated using \texttt{starry} \citep{luger2019starry}. This shows how an $\ell_{\mathrm{max}}=2$ representation fails to capture the spatial structure of the map; its brightest point is on the equator $24^{\circ}$ east of the substellar point, compared to the position of the brightest point in the original map at $40^{\circ}$ east. This offset is due to the low-order representation omitting many of the higher-order structures in the true map \citep{hammond2024wasp43b}. 

The $\ell_{\mathrm{max}}=4$ representation in Figure \ref{fig:gcm_true_l2_l4} captures most of the spatial structure, with its brightest point at $43^{\circ}$ east, more closely matching the brightest point of the original map. This implies that fitting an eclipse map composed of $\ell_{\mathrm{max}}=2$ harmonics would not produce an accurate fit to the real structure, while the $\ell_{\mathrm{max}}=4$ representation would in theory be able to match the real structure closely. For real observations where we do not know the spatial structure in advance, it is desirable to use a high degree of spatial freedom, constrained only by observational precision.

To simulate the light curve with \texttt{starry} \citep{luger2019starry}, we use the full map represented precisely by harmonics up to $\ell_{\mathrm{max}}=10$. WASP-43b is assumed to be tidally locked to its star, based on its short orbital period and consistent thermal phase curve amplitude, \citep{hellier2011wasp43b} so the fitted eclipse map corresponds to its permanent dayside. The simulated time series runs from orbital phase -0.6 to 0.4, covering a single orbit with an eclipse and a transit. It has 7000 data points, giving a cadence of 10.04 seconds per point, similar to the cadence of the JWST observations of WASP-43b in \citet{bell2024nightside}. 

We add Gaussian noise with a standard deviation $\sigma$, testing data with $\sigma= $ 150 ppm,  250 ppm, and  2000 ppm. We chose the value of 250 ppm to be comparable to the precision achieved by averaging the two eclipses of WASP-43b observed with JWST MIRI/LRS in \citet{bell2024nightside} and mapped in \citet{hammond2024wasp43b}. The value of 150 ppm represents a best-case scenario that could be achieved by averaging five such eclipses. The value of 2000 ppm represents low-precision data where there is no discernible mapping signal in the eclipse shape, but the rest of the phase curve can be resolved.

\subsection{Observed time-series thermal emission}\label{sec:methods:observations}

We also apply our method to three different observational datasets, plotted later in Section \ref{sec:results_obs}. They are all time-series observations of thermal emission from hot Jupiters. The first is a JWST MIRI/LRS observation of WASP-43b, averaged from 5 to 10.5 $\mu$m \citep{bell2024nightside}. The second is a JWST NIRISS/SOSS observation of WASP-18b, averaged from 0.85 to 2.85 $\mu$m \citep{coulombe2023broadband}. The final is a Spitzer Space Telescope observation of HD 189733b at 8 $\mu$m, compiled from seven separate eclipse observations \citep{majeau2012two}. Although the JWST datasets contain spectroscopic information, we average over all available wavelengths to achieve the highest possible precision.

A systematic model of instrumental effects and an astrophysical model of the system parameters was fitted to each dataset in each of the original studies to produce the final light curves \citep{majeau2012two,coulombe2023broadband,hammond2024wasp43b}. In this study, we do not explore the possibility of degeneracies between the systematic model, the astrophysical model, and the map model that we fit. Instead, we assume that the first two models are correctly known in advance, and only fit for the map (or a Fourier series model). In reality, there may be degeneracies between these models \citep{de2012towards,hammond2024wasp43b}. The JWST MIRI/LRS light curve we use for WASP-43b had its systematic and astrophysical model fitted alongside an $\ell_{\mathrm{max}}=2$ eclipse map, which should at least partially mitigate such degeneracies \citep{hammond2024wasp43b}.

\subsection{Fitting thermal emission with a Fourier series}\label{sec:methods:fourier}

The simplest model of a light curve of a planetary system uses a Fourier series for the variation in flux with time, multiplied by the eclipse and transit shapes of a planet emitting uniformly from its surface. We refer to this as a ``Fourier series model'' with flux $F(t)$:
\begin{multline}
    F(t) =  \biggl( A_{0} + \sum_{n=1}^{n_{\mathrm{max}}}  \biggl[ A_{n}\sin(2\pi n (t-t_{0})/P) \\ + B_{n}\cos(2\pi n (t-t_{0})/P)  \biggr]  \bigg) F_{p,0}(t) + F_{s}(t), 
\end{multline}
where $t_{0}$ is the transit time, $P$ is the orbital period, $A_{n}$ and $B_{n}$ are the fitted parameters up to order $n_\mathrm{max}$, $F_{p,0}$ is the time-series flux from a planet with uniform emission, and $F_{s}$ is the stellar flux including the transit model. We refer to a Fourier series model by its maximum order, so a model with maximum order $n_{\mathrm{max}}=2$ is $\mathcal{M}(n_{\mathrm{max}}=2)$.

This fits the out-of-eclipse light curve as well as any eclipse mapping model. However, it does not represent the effect of non-uniform planetary flux on the eclipse ingress and egress shapes. Strictly speaking, the eclipse shape in this model is slightly modified from the eclipse shape of a uniform planet by the small change in the Fourier series model over the short duration of the eclipse.

We will use this Fourier series as our null hypothesis when fitting datasets. To justify fitting an eclipse map, the map must fit the dataset better than this null hypothesis. Crucially, the map must not achieve this better fit by overfitting to noise in the eclipse shape and must actually be a better model of the true eclipse shape. We use an $\mathcal{M}(n_{\mathrm{max}}=2)$ Fourier series model as our null hypothesis as this is sufficient to capture almost all the out-of-eclipse variation of a phase curve, due to the rapidly decreasing contribution to a light curve from higher-order modes \citep{cowan2008inverting}.

\subsection{Fitting eclipse maps}\label{sec:methods:mapping}

For data of sufficient precision, we can fit an eclipse map model of the two-dimensional flux from the dayside of the planet using the information encoded in the shape of the eclipse. We refer to an ``eclipse map model'' by its maximum spherical harmonic degree $\ell_{\mathrm{max}}$, so an eclipse map model with maximum degree $\ell_{\mathrm{max}}=2$ is referred to as $\mathcal{M}(\ell_{\mathrm{max}}=2)$.

We define the map in terms of pixels evenly spaced in planetary surface area, known as a Mollweide projection \citep{snyder1987map}. These pixels uniquely define a spherical harmonic representation which continuously covers the mapped surface. \texttt{starry} uses this spherical harmonic representation to calculate the time-series emission from the map. Two-dimensional information is only accessible in the part of the surface eclipsed in both ingress and egress; this approximately corresponds to the dayside, corrected slightly for planetary rotation during the eclipse. We simplify this by plotting the posterior distribution of the fitted map on the dayside only, and plot the median fitted map globally for completeness.

The pixel representation and spherical harmonic representation of the map are interchangeable. Sampling the pixels makes it easier to require positive emission from the planet, as we simply place positive log-normal priors on the brightness of each pixel. These have a mean value of the peak brightness of the lightcurve divided by $\pi$, and a scale factor of 0.1 to set a very wide prior spanning multiple orders of magnitude in planetary flux (see the documentation of PyMC3 for the precise form of this prior \citep{exoplanet:pymc3}). This enforces positivity at the pixel locations only, so sometimes leads to the fitting of small areas of negative emission between pixels. 

If we sampled the spherical harmonics directly instead, we could not impose positivity as a prior, and would need to measure the positivity of the map at each step while it is sampled, which we found to be computationally intensive. We also found qualitatively that sampling pixels detected small-scale features more effectively than sampling the spherical harmonics. We suggest that this could be because each parameter in the pixel representation corresponds directly to a spatial region, so a small localised feature can be added by varying one parameter only (or a small number of parameters). On the other hand, introducing a new small-scale feature when sampling the spherical harmonic basis may require the sampler to adjust the values of all the spherical harmonic coefficients, as each one affects the entire map. The dense mass matrix approach available in PyMC3 \citep{exoplanet:pymc3} (first implemented in \citet{exoplanet:exoplanet}) allows us to efficiently sample the degeneracies in the posterior distribution of the brightness of the pixels.

The number of mapped pixels is defined by the maximum spherical harmonic order, increased by an oversampling factor that adds additional pixels to better cover the planetary surface. Increasing the spherical harmonic order increases the complexity of the features represented by the modelled light curve, while increasing the oversampling factor improves the accuracy of the measurement of map entropy in Section \ref{sec:methods:entropy}. Increasing both of these factors is desirable but requires more computation time. 

For our high-order maps, designed to have high spatial freedom, we found that a spherical harmonic basis of $\ell_{\mathrm{max}}=4$ and an oversampling factor of 3 (resulting in 62 pixels to sample) produced a good compromise between spatial freedom and computational time. For our low-order maps, designed to compare directly to the null hypothesis of an $n_{\mathrm{max}}=2$ Fourier series, we use a basis of $\ell_{\mathrm{max}}=2$ and an oversampling factor of 3 (corresponding to 16 pixels). The largest-scale eclipse mapping signals may be approximately fitted with an $\ell_{\mathrm{max}}=1$ map, but such a map would not in general be capable of fitting the phase curve outside eclipse. This is because the $n=2$ modes are still significant for the phase curve of a generic map \citep{cowan2008inverting}, so we choose to use the $\ell_{\mathrm{max}}=2$ map to capture these.

We define the pixels for a given map as a vector of brightness values $\mathbf{p}$. The sampled pixel values $\mathbf{p}$ define the vector $\mathbf{c}$ of the coefficients $c_{i}$ of the spherical harmonic representation via a matrix $\mathbf{M}$:
\begin{equation}\label{eqn:pixel_transform}
\mathbf{c} = \mathbf{M} \cdot \mathbf{p}.
\end{equation}
The matrix $\mathbf{M}$ is pre-calculated for our method by \texttt{starry} as it is unique to each combination of spherical harmonic order $\ell_{\mathrm{max}}$ and degree of oversampling. The 2D eclipse map $Z(\theta,\phi)$ is then constructed as a sum of the spherical harmonic basis maps $z_{i}(\theta,\phi)$ (where $\theta$ and $\phi$ are longitude and latitude), weighted by the coefficients $c_{i}$ of the vector $\mathbf{c}$:
\begin{equation}
Z(\theta,\phi) = \sum_{i} c_{i} z_{i}(\theta,\phi).
\end{equation}

The time-series emission $F(t)$ from the map defined by the coefficients $c_{i}$ is then a sum of the time-series emission $f_{i}(t)$ from each individual harmonic $z_{i}(\theta,\phi)$ (see \citet{luger2019starry} for their forms) weighted by its corresponding coefficient:
\begin{equation}\label{eqn:fit-curve}
F(t) = \sum_{i} c_{i} f_{i}(t).
\end{equation}
The modelled light curve $F(t)$ is therefore ultimately a function of the vector of the pixel brightness values $\mathbf{p}$, which are the parameters that we sample. 

We fit a model $\mathcal{M}$ to the data points $D$ in the light curve using PyMC3 \citep{exoplanet:pymc3}. We sample the posterior distribution of the brightness values of the pixels $\mathbf{p}$ with 250 samples in 8 chains, thinned by a factor of 4. This ensured that the Gelman-Rubin statistic was below 1.1 for our fitted maps. It would be possible to fit the orbital and instrumental parameters at the same time as this map \citep{hammond2024wasp43b}, but in this study we assume these parameters are known in advance. The sampling process for one map took around 15 minutes for an $\ell_{\mathrm{max}}=4$ map with 62 pixels using 8 cores on a recent Intel-based HPC server. Calculating the optimal information content of the map with cross-validation requires re-fitting the model around 300 times, resulting in a total runtime of several days. We settled on this number of samples as we did not find a noticeable improvement in mapping performance with more samples.

The $\ell_{\mathrm{max}}=4$ is intended to be the highest-order map that is computationally possible, as we aim for the most degrees of spatial freedom. This is why we do not consider $\ell_{\mathrm{max}}=3$ maps, as we only need a ``simplest model'' ($\ell_{\mathrm{max}}=2$) to detect a signal, and then a ``most complex model'' ($\ell_{\mathrm{max}}=4$) to fit the most accurate map. We chose the $\ell_{\mathrm{max}}=4$ map based on the sampling time, as using $\ell_{\mathrm{max}}=5$ maps took several times the sampling time, and sometimes did not sample the posterior fully. Using $\ell_{\mathrm{max}}=5$ maps also produced no noticeable improvement in the optimised map as any small-scale variation due to the $\ell=5$ modes was smoothed out when the information content was optimised.

\subsection{Model comparison with AIC and BIC}\label{sec:methods:aic_bic}

Eclipse maps with more parameters (more surface pixels or a higher spherical harmonic order) can fit smaller-scale features. This lets them fit an observed time series more closely, but they may use this freedom to overfit to the observational noise, mapping spurious small-scale features that do not really exist and increasing the uncertainty of the map. On the other hand, maps with too few parameters may not be able to fit the observed eclipse shape. Comparing models by their likelihood will always prefer more complex models that may overfit the data. We therefore need statistical metrics that penalise using too many parameters to overfit the data.

The Bayesian Information Criterion (BIC) \citep{schwarz1978estimating} and the Akaike Information Criterion (AIC) \citep{akaike1981likelihood} are metrics that address this issue, as they penalise the model likelihood by a function of the number of model parameters. Previous studies have used the BIC to address issues with eclipse map models overfitting small-scale features to eclipse shapes \citep{rauscher2018more,mansfield2020eigenspectra,challener2022theresa}. The BIC is:
\begin{equation}
    \mathrm{BIC}=k \ln (N)- 2 \ln p,
\end{equation}
where $k$ is the number of model parameters, $N$ is the number of data points, and $p$ is the model likelihood:
\begin{equation}
    \ln p = -\frac{1}{2} \sum_{i}^{N} \left( \frac{\mathcal{M}_{i} - D_{i}}{\sigma} \right)^{2} - N \ln \sigma -\frac{N}{2}\ln (2\pi),
\end{equation}
for a sum over all $N$ data points $D_{i}$ with uncertainty $\sigma$, fitted by a model $\mathcal{M}$ with values $\mathcal{M}_{i}$. A smaller BIC implies a better model, either through a good fit to the data or a small number of parameters. The BIC allows comparison of nested models where it is assumed that the true model is inside the set of tested models \citep{burnham2004multimodel}. We also consider the AIC of each model, which is:
\begin{equation}
    \mathrm{AIC}= 2 k - 2 \ln p.
\end{equation}
Unlike the BIC, the AIC assumes that the true model is only approximated by the set of tested models \citep{burnham2004multimodel}. This is essentially a philosophical difference to the BIC, and there is not a clear answer about which metric is more appropriate for our case. This is part of our motivation for using the data-driven cross-validation approach instead.

\subsection{Model comparison with cross-validation}\label{sec:methods:cv}

We propose the use of k-fold cross-validation to compare models of time-series thermal emission from exoplanet eclipses, as it penalises both underfitting and overfitting. Cross-validation is a metric of the ability of a model to predict data outside the dataset it is fitted to. In our case, the method removes a section of observed data  with size $k$, re-fits the model, and tests how well the re-fitted model predicts the omitted data \citep{hastie2009elements}. We found that k-fold cross-validation was better suited to tuning a regularisation parameter (details of which are below), compared to leave-one-out cross-validation \citep{challener2023eclipse}. We suggest this is because omitting a k-fold (with a longer duration than a single point) produces a stronger test of the predictive ability of a model due to the local autocorrelation in the time-series data.

There is no objectively correct way to choose the size of a k-fold, although $10\%$ of the size of a dataset is often used  \citep{hastie2009elements}. We chose a k-fold size of $20\%$ of the size of the ingress of an eclipse (i.e. $10\%$ of the total dataset containing two-dimensional information) which we found to be a good test of the predictive ability of a model. For the typical cadence and eclipse duration of our datasets, this results in k-folds containing roughly 10 data points each, in line with \citet{arlot2016choice}.

This choice of the width of the k-folds will limit the size of the spatial features that can be resolved with our mapping method. Each fold omits a stripe approximately  $36^{\circ}$ wide from either ingress or egress, and tests the ability of the model to predict the brightness of this stripe. This width is not a hard limit in itself, as the stripe is still measured across a different angle by the other ingress or egress, and the model still has access to the magnitude and gradient of the brightness either side of the omitted stripe. However, it must places some upper limit on the precision of the map and so this choice could be varied in future work.

Measuring the k-fold cross-validation score is a time-consuming process as the eclipse map must be re-fitted every time a new k-fold is removed. We greatly speed up this process by only testing k-folds in the ingress and egress of the eclipses, as well as sections of the light curve on either side of the eclipse with durations equal to the eclipse ingress and egress. This omits most of the out-of-eclipse phase curve from the test, but we found that this makes no practical difference. For typical datasets, the model only fails to fit the phase cure if it has far too few degrees of freedom, which our small test set easily identifies. For datasets with multiple eclipses, we apply these k-folds periodically, removing the same section of each eclipse (or either side of it). This is the same as stacking the eclipses (averaging them together) and removing the k-fold from the averaged eclipse. We already effectively stack the eclipses by fitting a periodic model that cannot vary from one eclipse to another.

We split the tested part of the dataset (the eclipse ingress and egress, and sections of the phase curve of equal duration either side of the eclipse) into $K=20$ sets, each of length $N$ points. When leaving out the $k^{\mathrm{th}}$ fold, we label the dataset as $D_{-k}$, and the model as $\mathcal{M}^{-k}$. The cross validation score for the $k^{\mathrm{th}}$ fold is:
\begin{equation}\label{eqn:cv_k}
CV^{k} = \frac{1}{N}\sum ^{N}_{i=0} \log \left( \frac{1}{S} \sum_{s=1}^{S} p(D_{i} | \mathcal{M}^{-k}) \right),
\end{equation}
where $p(D_{i} | \mathcal{M}^{-k})$ is the likelihood of each point $D_{i}$ in the k-fold, given the model $\mathcal{M}^{-k}$ fitted to the dataset $D_{-k}$. For each data point, we sum over the $S$ samples drawn from the posterior distribution as in \citet{vehtari2017practical}, and then sum over each of the $N$ data points in the $k^{\mathrm{th}}$ fold. $CV^{k}$ therefore tests the ability of the model to predict the points in the $k^{\mathrm{th}}$ fold when it is fitted without access to these points. In the context of eclipse mapping, it tests the ability of a model to predict the brightness of a small slice across the planetary disk when the contribution of that slice to the eclipse shape is removed. When we use this method to optimise a fitted map, we always plot the map fitted to the full dataset and never explicitly plot maps fitted to datasets with data removed from k-folds.

We find the overall cross-validation score $CV_{\mathcal{M}}$ for a model $\mathcal{M}$ by averaging over all $K$ folds:
\begin{equation}
CV_{\mathcal{M}} = \frac{1}{K}\sum_{k=1}^{K} CV^{k}.
\end{equation}
The cross-validation score penalises underfitting (when the model cannot fit the data) as the predictive likelihood $p(D_{i} | \mathcal{M}^{-k})$ is low for a poorly fitting model. Importantly, it also penalises overfitting because a model $\mathcal{M}^{-k}$ with too many degrees of freedom will produce a posterior distribution in the $k^{\mathrm{th}}$ fold that is much wider than the true spread of the omitted data $D_{k}$. This means that the average predictive likelihood $p(D_{i} | \mathcal{M}^{-k})$ will be low when summed over all $S$ samples, resulting in a low cross-validation score.

Conversely, a well-fitting model will predict the omitted datapoints accurately and precisely as it is fitting the underlying physical system (instead of overfitting to noise), so the data points in one particular k-fold are implied by those in the others. This will result in a model $\mathcal{M}^{-k}$ with a posterior distribution with a similar mean and standard deviation to the real omitted data $D_{k}$, producing a high predictive likelihood $p(D_{i} | \mathcal{M}^{-k})$ and a high cross-validation score.

To compare the cross-validation score for two competing models $\mathcal{M}_{1}$ and $\mathcal{M}_{2}$, we take the difference between them:
\begin{equation}\label{eqn:delta_cv}
\Delta CV_{\mathcal{M}_{1}\mathcal{M}_{2}} = CV_{\mathcal{M}_{1}} - CV_{\mathcal{M}_{2}}.
\end{equation}
The standard deviation of this estimate is calculated from the variance $V$ of the difference in the scores in each k-fold \citep{chen2021one,welbanks2023application}:
\begin{equation}
\mathrm{SD}(\Delta CV_{\mathcal{M}_{1}\mathcal{M}_{2}}) = \sqrt{V_{k=1}^{K} (CV^{k}_{\mathcal{M}_{1}} - CV^{k}_{\mathcal{M}_{2}})}
\end{equation}
where $CV^{k}_{\mathcal{M}_{1}}$ and $CV^{k}_{\mathcal{M}_{1}}$ are calculated from Equation \ref{eqn:cv_k} for each individual k-fold.  The standard error is then calculated from the standard deviation \citep{chen2021one}:
\begin{equation}\label{eqn:se_delta_cv}
    \mathrm{SE}(\Delta CV_{\mathcal{M}_{1}\mathcal{M}_{2}}) = \frac{\mathrm{SD}(\Delta CV_{\mathcal{M}_{1}\mathcal{M}_{2}})} {\sqrt{K}}
\end{equation}
Due to the assumptions made in deriving this standard error estimate, we caution against using it as an ``multiple-sigma'' preference for one model over another, and restrict its use to identifying a simple preference one way or the other. The standard error of the difference in cross-validation score between two models tells us if the difference is significant or not. If two models have consistent cross-validation scores within this standard error estimate, we assess that they are as good as each other (and therefore we prefer the simpler model). 

The exception to this is when we later tune a regularisation parameter to find the best cross-validation score; in that case, we select the model with the highest cross-validation score as the best-performing, even though other nearby values of the regularisation parameter will necessarily be within one standard error. We chose not to use the the ``one standard error'' rule \citep{chen2021one} in this case, as we have no \textit{a priori} tendency to prefer a higher or lower regularisation parameter. 

\subsection{Optimising the spatial scale of fitted maps}\label{sec:methods:entropy}

The eclipse maps fitted with the method in Section \ref{sec:methods:mapping} have a spatial scale determined by their number of pixels (corrected for oversampling). Maps with a small number of pixels (or low-order spherical harmonics) can only fit very large-scale features with similarities to the shapes of the low-order spherical harmonics. Figure \ref{fig:gcm_true_l2_l4} shows how a low-order spherical harmonic representation may not be able to fit the true shape of a map. Maps with large numbers of pixels (or high-order spherical harmonics) can fit spatial features more accurately, as shown in Figure \ref{fig:gcm_true_l2_l4}. However, they may also produce spurious small-scale features due to the noise and degeneracy in the observed light curve. To address this issue, we use a large number of pixels to allow freedom in fitting small-scale features, but apply the method of \citet{horne1985images} (applied to brown dwarfs in \citet{chen2024map}) to select maps with an appropriate spatial scale given the precision of the data. 

\begin{figure*}
\centering 
  \includegraphics[width=\textwidth]{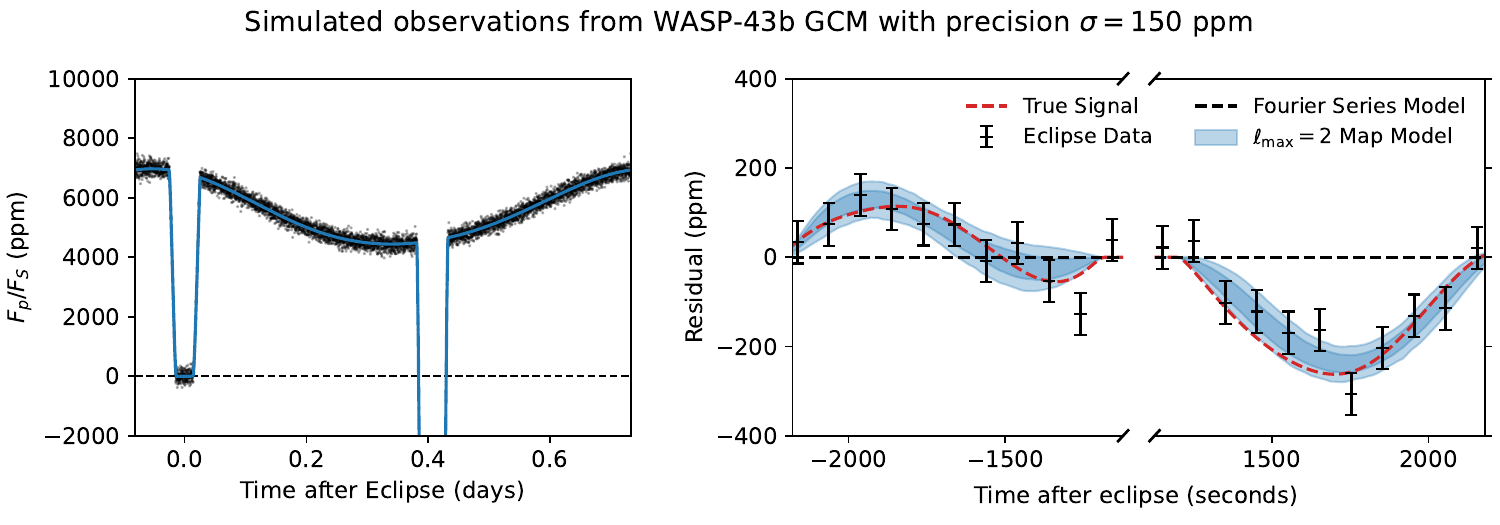}%
  
  \includegraphics[width=\textwidth]{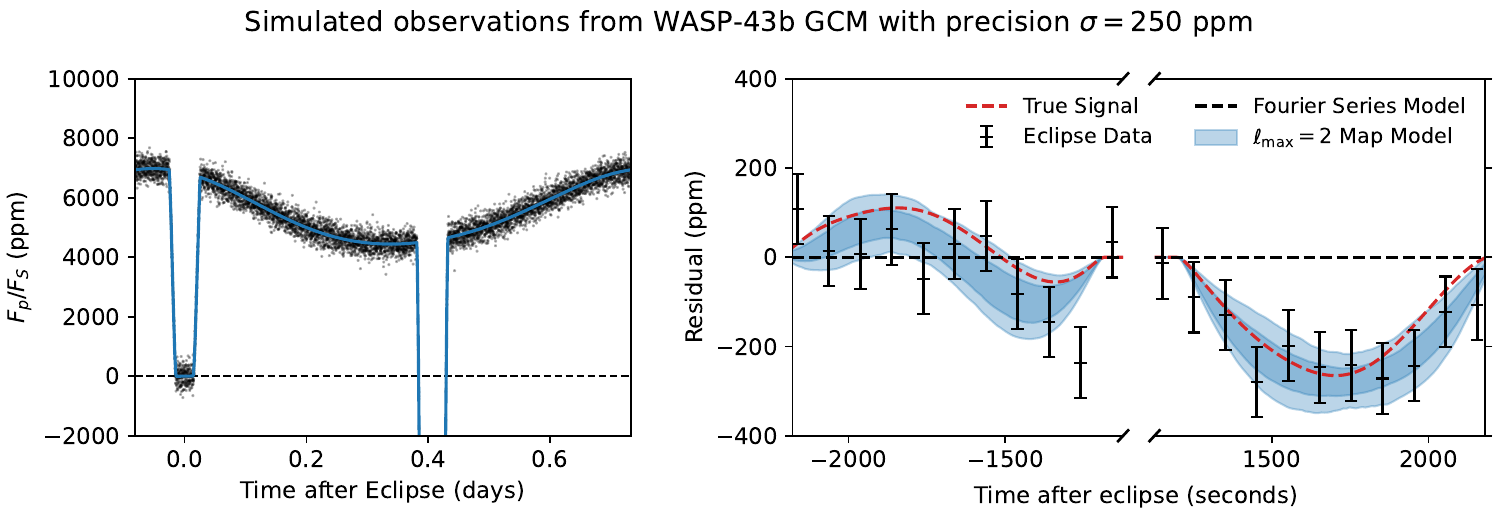}%
\caption{Two simulated datasets with clear eclipse mapping signals. First row: a simulated observation (with precision $\sigma=150$ ppm) of time-series thermal emission from the simulation of WASP-43b in the \texttt{THOR} GCM. The left-hand panel shows the dataset fitted by an eclipse map model $\mathcal{M}(\ell_{\mathrm{max}}=2)$. The right-hand panel shows the residual eclipse mapping signal, which is the data (binned every 10 points) subtracted by a fit with a Fourier series model $\mathcal{M}(n_{\mathrm{max}}=2)$ that assumes a uniform planetary disk. There is a clear residual signal (dashed red line) that the data (black points) resolves, which the eclipse map model (blue region) fits better than the Fourier series model (dashed black line). The blue region showing the map model contains two shaded regions showing the first and second quantiles, containing 68.27\% and 95.45\% of the posterior distribution. Second row: a simulated observation (with precision $\sigma=250$ ppm) of time-series thermal emission from the simulation of WASP-43b in the \texttt{THOR} GCM. As in the first row, there is a clear eclipse mapping signal which the eclipse mapping model fits better than the Fourier series model, although with slightly less certainty due to the lower-precision data. }\label{fig:sim_w43b_good}
\end{figure*}

We measure the information content of a map from the spatial entropy of its pixels \citep{gull1978image}. We seek to match the information content of the map to the information content of the dataset. A map with more information than the dataset will overfit, and a map with less information than the dataset will underfit. We follow \citet{horne1985images} and consider the entropy $S$ of the pixels defining a map $I$ composed of pixels denoted by $j$:
\begin{equation}
    S=-\sum_{j} I(j) \ln \left[\frac{I(j)}{D(j)}\right] \text {. }
\end{equation}
where $D$ is the ``default image'' defined by \citet{horne1985images}. This default image depends on a weighting function $w(k,j)$ which defines the relation of all pixels $j$ to an individual pixel $k$:
\begin{equation}
    D(k)=\frac{\sum_{j} w(k, j)\ I(j)} { \sum_{j} w(k, j)}.
\end{equation}
We use a weighting function $w(k,j)=1$ so that the entropy tracks deviations from a uniform map. This matches our null hypothesis, which is the eclipse shape of a uniformly bright planetary disk as defined in Section \ref{sec:methods:fourier}. 

We include the information content of the map in the fitting process by penalising the model likelihood $p$ by a factor $2\alpha S$. This provides a functional that balances the model likelihood against the mapped complexity, with the balance set by the free parameter $\alpha$, as originally used in \citet{vogt1987doppler}. We follow the notation in \citet{chen2024map} and formulate the functional as the likelihood penalised by the entropy:
\begin{equation}\label{eqn:L_penalty}
    \hat{p} = p - 2\alpha S,
\end{equation}
where $\alpha$ is a regularisation parameter. A large value of $\alpha$ promotes maps with low entropy and large-scale features, which may underfit the data.  A small value of $\alpha$ allows maps with high entropy and small-scale features, which may overfit the data. We aim to find an optimal value of $\alpha$, which achieves the best k-fold cross-validation score. We therefore vary $\alpha$ until we find the value that results in the best cross-validation score. We expect that this will produce a map with an appropriate information content given the precision of the data. This method is analogous to fitting a spline to a one- or two-dimensional dataset, where there is a single regularisation parameter that gives an optimal cross-validation score \citep{wahba1990spline}.

\section{Simulated Results}\label{sec:results_sim}

In this section, we apply the methods from Section \ref{sec:methods} to simulated data. We first demonstrate the process of identifying eclipse mapping signals in simulated data using cross-validation. We then show how fitting an eclipse map with spherical harmonics alone can lead to under- or over-fitting. To remedy this, we show how we can optimise the information content (smoothness) of a high-order spherical harmonic map using cross-validation, and derive an optimal map for this simulated dataset. Note that in Sections \ref{sec:results_sim:sim_signal} and \ref{sec:results_sim:model_comparison_aic_bic_cv} we do not apply a penalty to the likelihood based on the entropy of the fitted map (effectively, we keep $\alpha=0$ until Section \ref{sec:results_sim:smooth_gcm}).

\subsection{Identifying eclipse mapping signals in simulated data}\label{sec:results_sim:sim_signal}

\begin{figure*}
\centering 
  \includegraphics[width=\textwidth]{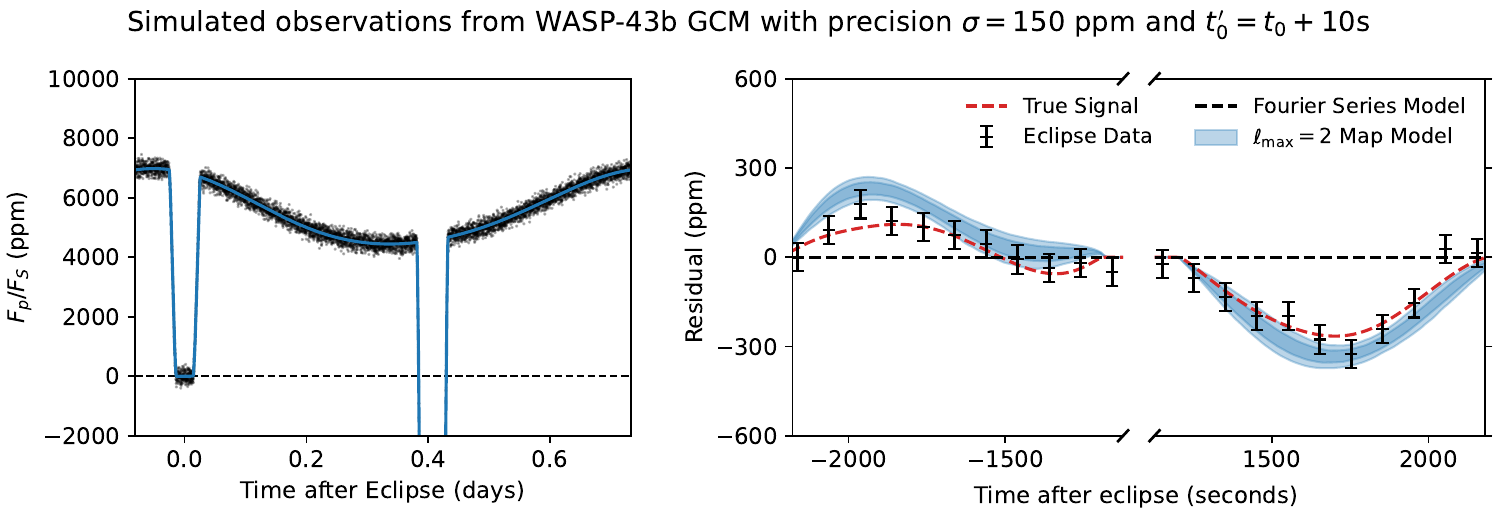}%
  
  \includegraphics[width=\textwidth]{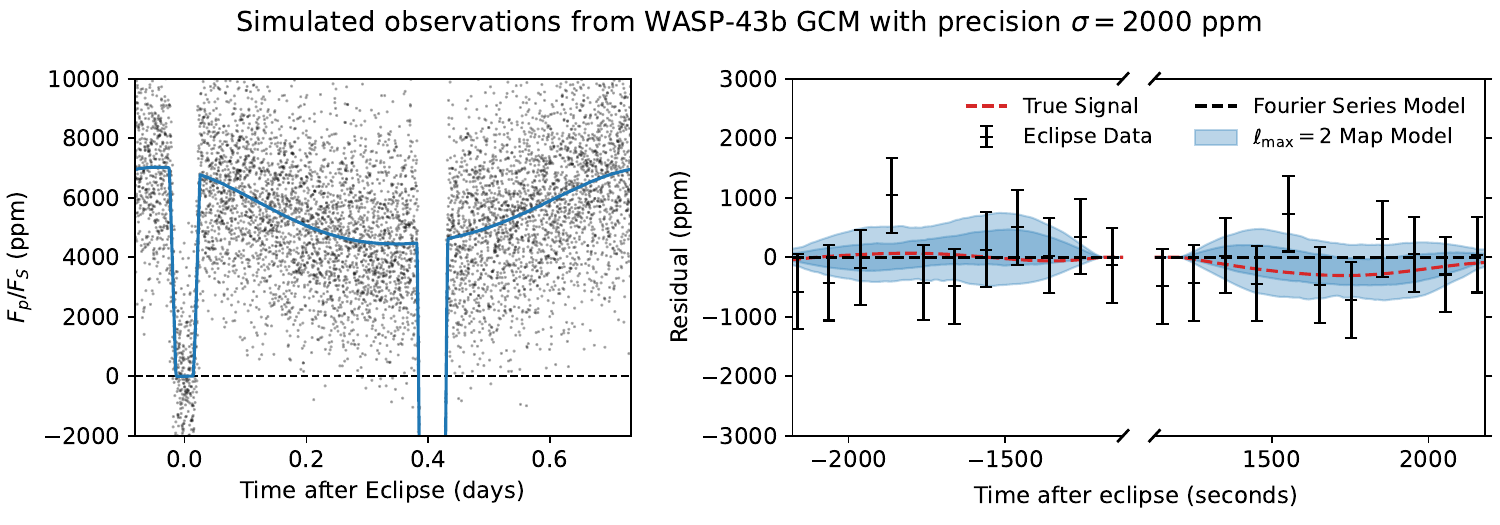}%
\caption{Two simulated datasets that are too inaccurate or too imprecise to fit an eclipse map. First row: a simulated observation (with precision $\sigma=150$ ppm) of time-series thermal emission from the simulation of WASP-43b in the \texttt{THOR} GCM, where the map is fitted with an incorrect value for the transit time of $t_{0}^{\prime}=t_{0}+10\mathrm{s}$. The layout of the panels is the same as in Figure \ref{fig:sim_w43b_good}. The incorrect astrophysical model means that the longitudinal structure imposed by the phase curve produces an incorrect fit to the data in the eclipse, shown by the mismatch between the black data points and the blue fitted model. Second row: a simulated observation (with precision $\sigma=2000$ ppm) of time-series thermal emission from the simulation of WASP-43b in the \texttt{THOR} GCM. The low precision of the data means that it does not resolve the eclipse mapping signal, so the $\ell_{\mathrm{max}}=2$ model does not fit this signal any better than the $n_{\mathrm{max}}=2$ Fourier series model (the dashed black zero line). The red dashed lines showing the true eclipse mapping signals are identical to those in Figure \ref{fig:sim_w43b_good}.}\label{fig:sim_w43b_bad}
\end{figure*}

As described in Section \ref{sec:methods:cv}, it is possible to fit spurious eclipse maps to datasets that are not accurate enough to support them. In this section, we therefore show how the cross-validation method described in Section \ref{sec:methods:cv} can quantitatively identify the presence or absence of an eclipse mapping signal in a dataset.

\begin{figure*}
\centering 
  \includegraphics[width=\textwidth]{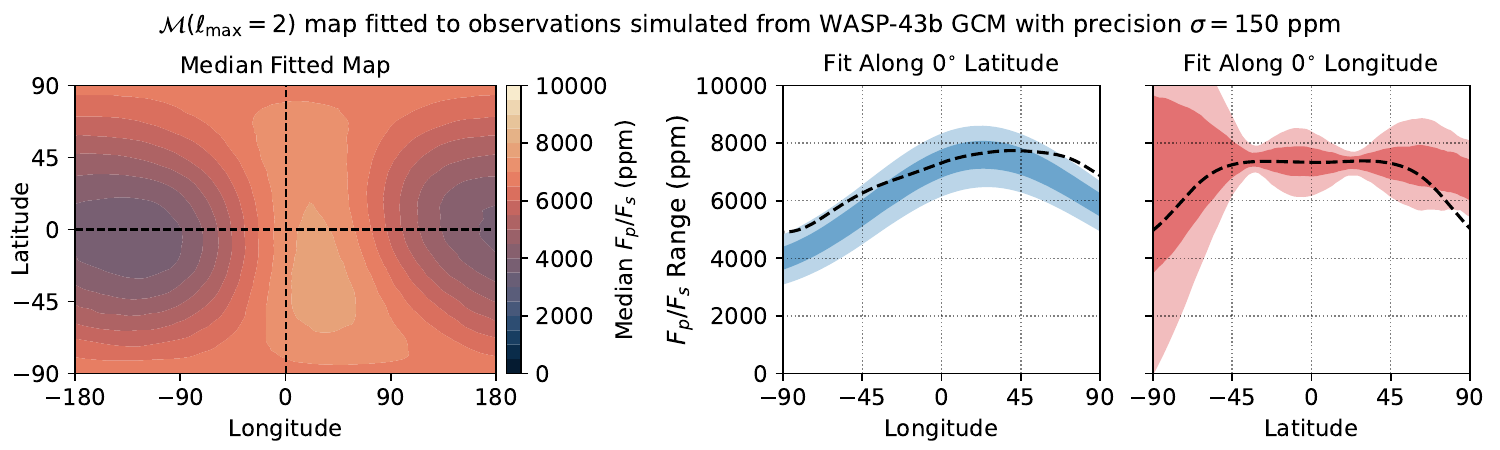}%
\caption{An eclipse map model $\mathcal{M}(\ell_{\mathrm{max}}=2)$ fitted to the 150 ppm simulated data in Figure \ref{fig:sim_w43b_good}. The first panel shows a map defined by the median values of the sampled distribution at every location. The second and third panels show the fitted map as a function of latitude and longitude through the substellar point, with two shaded regions showing the first and second quantiles, containing 68.27\% and 95.45\% of the posterior distribution. The limited freedom of the $\ell_{\mathrm{max}}=2$ map means that it cannot exactly fit the true map from the GCM simulation (the dashed black lines, shown in full in Figure \ref{fig:gcm_true_l2_l4}) which has more complex structures composed of higher-order harmonics. This results in the brightest point on the equator being measured at $(24^{+0}_{-0})^{\circ}$ east, compared to the true value of $40^{\circ}$ east shown in Figure \ref{fig:gcm_true_l2_l4}. The lack of uncertainty on this measurement is due to the strong constraint from the full phase curve on the longitudinal structure of the fitted low-order spherical harmonics.}\label{fig:sim_gcm_l2_150}
\end{figure*}

\begin{table*}
    \centering
    \begin{tabular}{lcccc}
         \textbf{Dataset} & $\Delta \chi^{2}$ & $\Delta$BIC & $\Delta$AIC & $\Delta$CV   \\
         \hline
         WASP-43b GCM Simulation ($\sigma=150$) &  +126.04 & +81.77  & +116.04 &  $+0.148 \pm 0.016$ \\
         WASP-43b GCM Simulation ($\sigma=250$) & +73.37  & +29.1 & +63.37  &  $+0.076 \pm 0.007$ \\
         \hline
         WASP-43b GCM Simulation ($\sigma=150$, $t_{0}^{\prime} = t_{0} + 10 \mathrm{s}$) & +54.79 & +10.52 & +44.79  & $-0.002 \pm 0.013$  \\
         WASP-43b GCM Simulation ($\sigma=2000$)  & -0.31 & -44.58 & -10.31  & $-0.004 \pm 0.001$  \\
    \end{tabular}
    \caption{Statistical metrics for the eclipse map models $\mathcal{M}(\ell_{\mathrm{max}}=2)$ fitted to the datasets in Figures \ref{fig:sim_w43b_good} and \ref{fig:sim_w43b_bad}, relative to the metrics for the Fourier series model $\mathcal{M}(n_{\mathrm{max}}=2)$ (the null hypothesis). The $\chi^{2}$ value is the lowest $\chi^{2}$ value in the posterior of the fitted model, which is also used to derive the BIC and the AIC. A positive value for a metric means that the eclipse mapping model is preferred over the Fourier series model.}
    \label{tab:stats_comparison_sim_w43b}
\end{table*}

\begin{figure*}
\centering 
  \includegraphics[width=\textwidth]{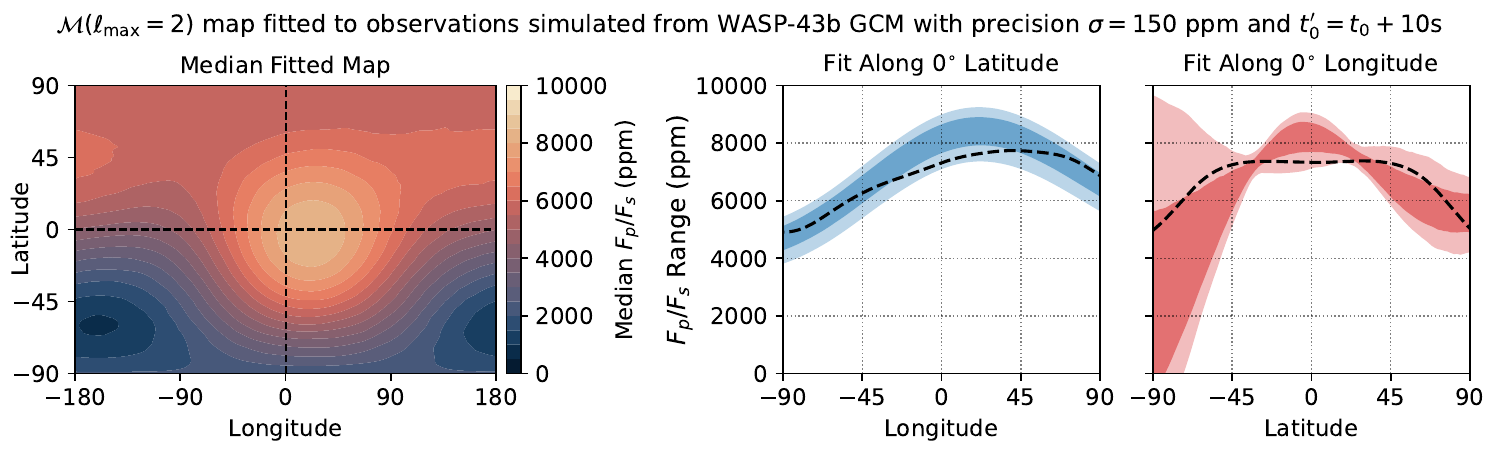}%
  
  \includegraphics[width=\textwidth]{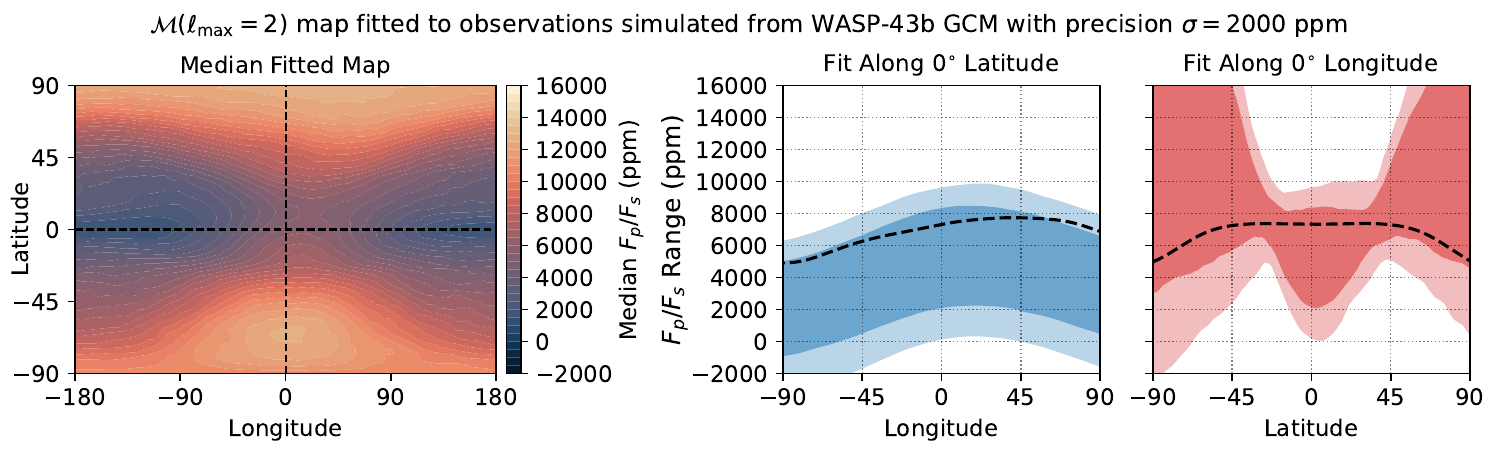}%
\caption{Eclipse map models $\mathcal{M}(\ell_{\mathrm{max}}=2)$ fitted to the simulated datasets in Figure \ref{fig:sim_w43b_bad}, with low accuracy and low precision respectively. First row: the map fitted to the inaccurate simulated data in the first row of Figure \ref{fig:sim_w43b_bad} with an incorrect transit time $t_{0}^{\prime}=t_{0}+10\mathrm{s}$. The layout of the panels is the same as Figure \ref{fig:sim_gcm_l2_150}. The systematic error on the eclipse timing means that the fitted map does not match the true map (the dashed black lines, shown in full in Figure \ref{fig:gcm_true_l2_l4}) but has spuriously high confidence. Second row: the map fitted to the imprecise data in the second row of Figure \ref{fig:sim_w43b_bad}. The fitted map has very high uncertainty but does include the true map inside its posterior distribution as it has no systematic errors like the error in eclipse time present in the map in the first row. Neither dataset is suitable for fitting an eclipse map, shown by the cross-validation score in Table \ref{tab:stats_comparison_sim_w43b} and Figure \ref{fig:cv_comparison_sim_w43b}.}\label{fig:sim_w43b_bad_maps}
\end{figure*}

Figures \ref{fig:sim_w43b_good} and \ref{fig:sim_w43b_bad} show light curves produced from the GCM simulation of WASP-43b as described in Section \ref{sec:methods:simulation}. Their left-hand panels show the time-series thermal emission. Their right-hand panels show the same data, binned every 10 points, and subtracted by a model of a planet with a uniform dayside emission. This leaves a residual signal which corresponds to the deviation of the eclipse shape from the eclipse shape of a uniform disk --- the ``eclipse mapping signal'', as discussed in \citet{hammond2024wasp43b}. A consistent deviation in the data points from the zero line implies an eclipse mapping signal. The dashed red lines show the true residual signal in eclipse before the Gaussian noise is added. A positive residual signal implies that the strip across the planetary disk covered by the stellar edge at that moment is brighter than a uniform disk producing the same total eclipse depth and vice versa. 

The first row of Figure \ref{fig:sim_w43b_good} shows the simulated data for $\sigma=$ 150 ppm. This represents a best-case scenario for observations with JWST, corresponding to an average of five eclipses of WASP-43b with MIRI/LRS \citep{bell2024nightside}. There is a clear eclipse mapping signal, shown by the large residual in the data in the right-hand panel, fitted well by the $\ell_{\mathrm{max}}=2$ eclipse map model.

Table \ref{tab:stats_comparison_sim_w43b} shows that the eclipse mapping model $\mathcal{M}(\ell_{\mathrm{max}}=2)$ achieves a significantly better $\chi^{2}$ value for the $\sigma=150$ ppm data than the Fourier series model $\mathcal{M}(n_{\mathrm{max}}=2)$, due to its better fit to the eclipse shape in the first row of Figure \ref{fig:sim_w43b_good}. This does not necessarily demonstrate the presence of two-dimensional information in the eclipse shape, as the eclipse mapping model has more degrees of freedom so may achieve a better $\chi^{2}$ score by overfitting to noise in the eclipse. Both models fit the out-of-eclipse phase curve as well as each other, as the $\ell_{\mathrm{max}}=2$ spherical harmonics produce the same out-of-eclipse phase curve components as the $n_{\mathrm{max}}=2$ Fourier components, apart from a small correction for inclination.

Figure \ref{fig:sim_gcm_l2_150} shows the eclipse map model $\mathcal{M}(\ell_{\mathrm{max}}=2)$ fitted to the $\sigma=$ 150 ppm dataset in Figure \ref{fig:sim_w43b_good}. The first panel shows the median of the posterior distribution of fitted maps (note that the two-dimensional structure is only constrained on the dayside). We plot the median at every point because we sample the pixels composing the map; this plot therefore shows the median value of the sampled posterior distribution. We confirmed (but do not plot) that this median map is very similar to the maximum likelihood map from the posterior distribution, as discussed in \citet{hammond2024wasp43b} This is the case in general for all of the maps in this paper with relatively narrow posterior distributions. This is not the case for high-order maps with very wide posterior distributions, where the maximum likelihood map can be very different to the median map, as it is overfitting to noise and derives a particular set of spurious small-scale features that happen to fit the noise best.

The next two panels of Figure \ref{fig:sim_gcm_l2_150} show slices of the posterior distribution of the fitted maps on the dayside through the substellar point, as functions of longitude and latitude. The true map is composed of many high-order spherical harmonic components beyond $\ell=2$, so this fitted $\ell_{\mathrm{max}}=2$ map cannot match the shape of the true map very closely. It does, at least, fit the residual signal accurately, so we suggest that this map is statistically supported, if not especially accurate. 

\citet{hammond2024wasp43b} considered the ''observable'' and ``null'' spaces of eclipse maps of GCM simulations as defined by \citet{challener2023eclipse}, which are the components of the map that do and do not contribute to the mapping signal respectively. We do not consider the observable space in this study as we want to instead highlight the loss of information from fitting a map with a specific set of spherical harmonics, as well as the ability of our optimisation method to access the true spatial scale of the GCM. However, it should be remembered that the black dashed lines showing the ``true'' GCM maps in Figure \ref{fig:sim_gcm_l2_150} and onwards are not exactly accessible by the conversion from light curve to spatial map; see \citet{challener2023eclipse} for more discussion.

The second row of Figure \ref{fig:sim_w43b_good} shows the simulated data for $\sigma=$ 250 ppm. This represents a typical scenario for observations with JWST, corresponding to an average of two eclipses of WASP-43b with MIRI/LRS \citep{bell2024nightside}. The large residual shows an eclipse mapping signal, although resolved less clearly than in the first row of Figure \ref{fig:sim_w43b_good}. Table \ref{tab:stats_comparison_sim_w43b} again shows that the eclipse mapping model $\mathcal{M}(\ell_{\mathrm{max}}=2)$ achieves a much better $\chi^{2}$ value than the Fourier series model $\mathcal{M}(n_{\mathrm{max}}=2)$ for this dataset.

The first row of Figure \ref{fig:sim_w43b_bad} shows the simulated data for $\sigma=$ 150 ppm, with the Fourier series and eclipse map models fitted using an incorrect value for the transit time $t_{0}^{\prime} = t_{0} + 10\mathrm{s}$. As we assume a circular orbit, this produces the same offset of $+10\mathrm{s}$ in the assumed eclipse time. This simple change produces a systematic error in the astrophysical model, which means that the eclipse mapping signal and the eclipse map are fitted poorly regardless of the precision of the data. An error in assumed eclipse time could also arise in other ways, such as a slightly eccentric orbit producing an offset in eclipse timing relative to a measured transit time. Fitting the model with this delayed eclipse timing results in an eclipse ingress that is too high, and an eclipse egress that is too low. The model wrongly adjusts the eclipse map to compensate for this error. \citet{williams2006resolving} showed how a timing error can produce a spurious longitudinal offset in a map from an isolated eclipse. We suggest that a longitudinal offset is not introduced in this case because the low-order longitudinal structure is strongly constrained by the out-of-eclipse phase curve (which is not significantly affected by the small timing error). The model therefore adjust the latitudinal structure instead to attempt to better fit the residual signal in the eclipse shape.

This example is intended to represent any source of systematic error in the fit of the model to the data, such as instrumental systematic effects or other incorrect orbital parameters. Both of these can produce significant deviations in the modelled eclipse shape \citep{de2012towards,hammond2024wasp43b}. The modelled eclipse map residual in the first row of Figure \ref{fig:sim_w43b_bad} fails to match the eclipse data accurately due to the systematic error in the astrophysical model caused by the incorrect eclipse timing. This is a distinctive signature of a map fitted with a systematic error --- the longitudinal structure imposed by the phase curve produces a residual shape that is not consistent with the observed residual; the latitudinal structure of the map is then wrongly adjusted to try to compensate for this error. The first row of Figure \ref{fig:sim_w43b_bad_maps} shows the eclipse map fitted to this dataset, which fails to match the true map, especially in its latitudinal structure.

Table \ref{tab:stats_comparison_sim_w43b} shows that the eclipse mapping model still achieves a better $\chi^{2}$ value than the Fourier series model for this dataset. This is despite the highly inaccurate map and despite the fact that its posterior distribution is highly statistically different from the true residual signal (compare the blue posterior distribution to the red dashed line in the first row of Figure \ref{fig:sim_w43b_bad}). The better $\chi^{2}$ value is due to the fact that the eclipse map model has increased freedom to fit the eclipse shape, even though this does not correctly correspond to a mapping signal.

\begin{figure*}
\centering 
  \includegraphics[width=\textwidth]{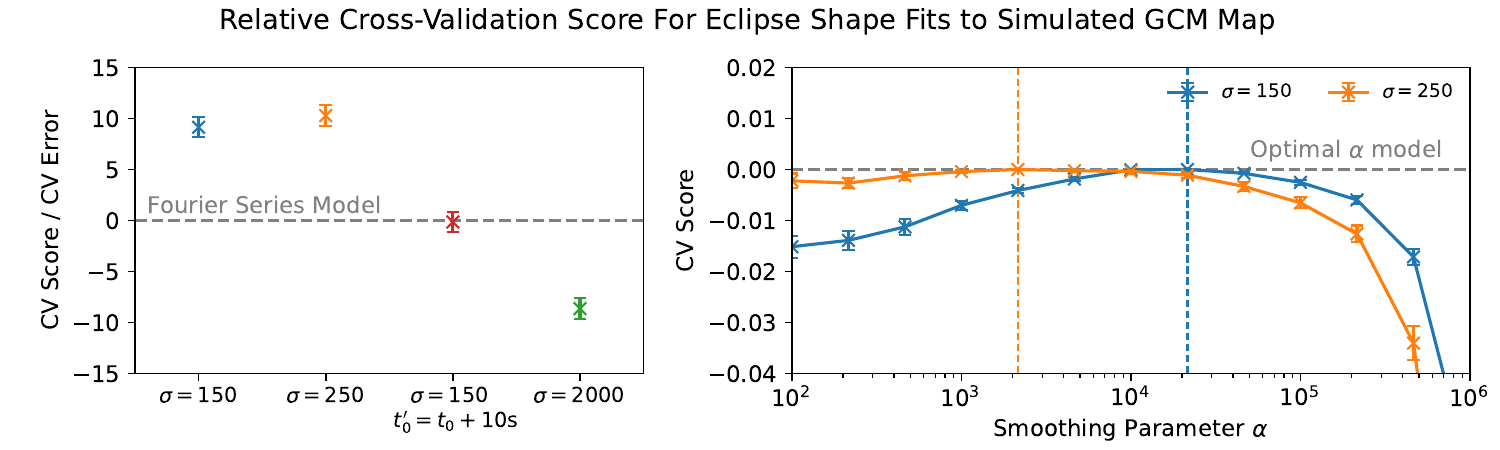}%
\caption{Left panel: The relative cross-validation scores (Equation \ref{eqn:delta_cv}) for eclipse map models $\mathcal{M}(\ell_{\mathrm{max}}=2)$ fitted to the simulated datasets in Figures \ref{fig:sim_w43b_good} and \ref{fig:sim_w43b_bad}, compared to the scores for Fourier series models $\mathcal{M}(n_{\mathrm{max}}=2)$. The scores are normalised by their standard errors (Equation \ref{eqn:se_delta_cv}). The eclipse map models $\mathcal{M}(\ell_{\mathrm{max}}=2)$ fitted to the higher precision data in Figure \ref{fig:sim_w43b_good} have significantly better cross-validation scores than the Fourier series models, indicating the presence of eclipse mapping signals. The eclipse map models $\mathcal{M}(\ell_{\mathrm{max}}=2)$ fitted to the inaccurate and imprecise data in Figure \ref{fig:sim_w43b_bad} achieve cross-validation scores that are lower than (or consistent with) the Fourier series model, indicating the absence of eclipse mapping signals. Right panel: the difference in cross-validation score for eclipse map models $\mathcal{M}(\alpha,\ell_{\mathrm{max}}=4)$ fitted to the simulated datasets in Figure \ref{fig:sim_w43b_good}, with their likelihood penalised by the regularisation parameter $\alpha$ as described in Section \ref{sec:methods:entropy}. This identifies an optimal information content (spatial scale) for the map in each case. The cross-validation scores are plotted relative to the optimal cross-validation score for each dataset, so the maximum value of each curve is zero.}\label{fig:cv_comparison_sim_w43b}
\end{figure*}

The second row of Figure \ref{fig:sim_w43b_bad} shows the simulated data for $\sigma=$ 2000 ppm. This represents a light curve that is too imprecise for eclipse mapping. The posterior distribution of the eclipse mapping model $\mathcal{M}(\ell_{\mathrm{max}}=2)$ does not match the true residual shape (the dashed red line). The second row of Figure \ref{fig:sim_w43b_bad_maps} shows the eclipse map fitted to this dataset, which is very uncertain due to the low precision of the data. Table \ref{tab:stats_comparison_sim_w43b} shows that the mapping model achieves almost the same $\chi^{2}$ value as the $\mathcal{M}(n_{\mathrm{max}}=2)$ model (actually a slightly worse value). Figure \ref{fig:sim_w43b_bad} shows how the two models fit the data essentially as well as each other as the precision of the data is too low to reveal the effect of the two-dimensional structure on the eclipse shape. 

These simulated datasets show different pitfalls that could be encountered when fitting eclipse maps. We need statistical metrics that will strongly penalise mapping models that use too many parameters to fit the eclipse shape of inaccurate or imprecise data.

\begin{figure*}
\centering 
 \includegraphics[width=\textwidth]{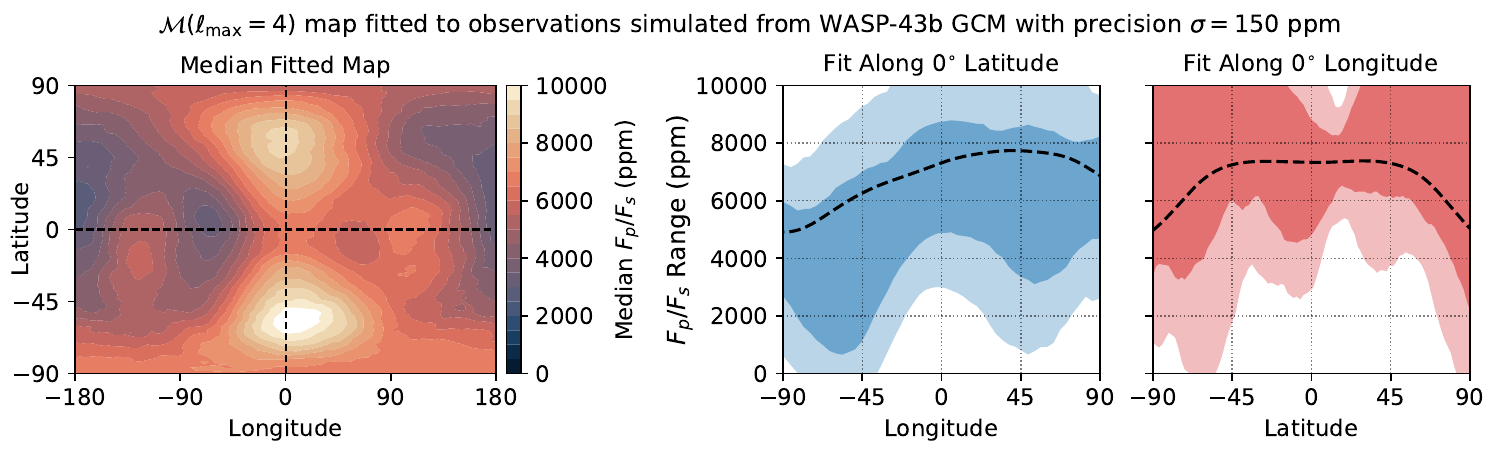}%
\caption{A high-order eclipse map model $\mathcal{M}(\ell_{\mathrm{max}}=4)$ fitted to the 150 ppm simulated data in Figure \ref{fig:sim_w43b_good}. The layout of the panels is the same as Figure \ref{fig:sim_gcm_l2_150}. The higher degree of spatial freedom means that it is possible in theory for the fitted map to closely match the true map. However, in practice the increased spatial freedom allows too large a range of solutions, making the uncertainty on the map very large. This makes the measurement of the brightest point on the equator very imprecise at $(56^{+65}_{-58})^{\circ}$ east, compared to the true value of $40^{\circ}$ east shown in Figure \ref{fig:gcm_true_l2_l4}.}\label{fig:sim_gcm_l4_150}
\end{figure*}

\begin{figure*}
\centering 
\includegraphics[width=\textwidth]{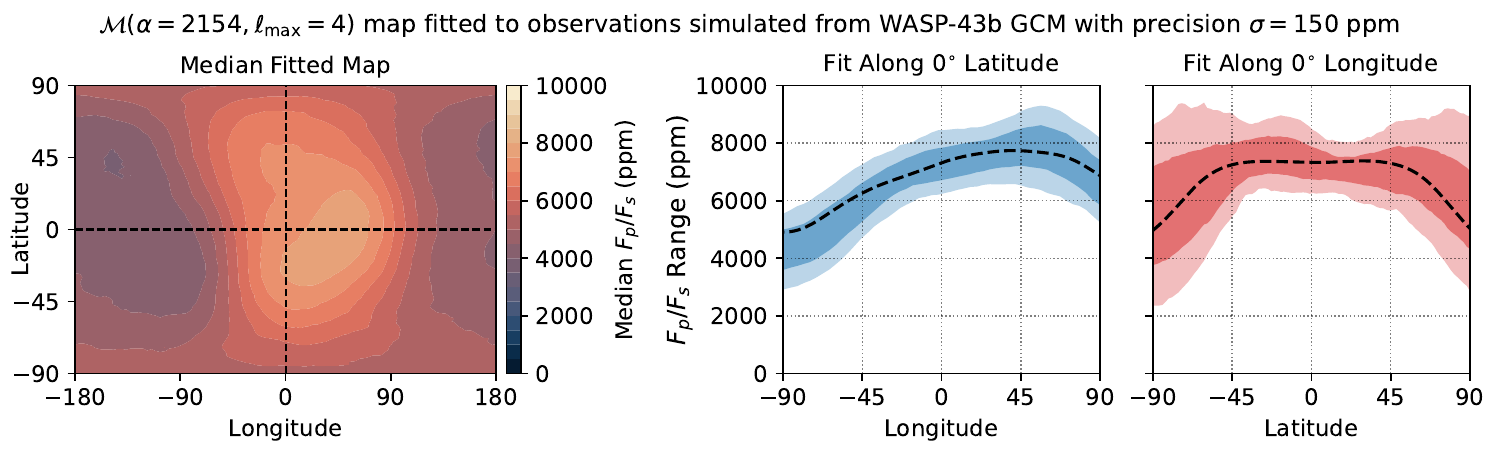}%

\includegraphics[width=\textwidth]{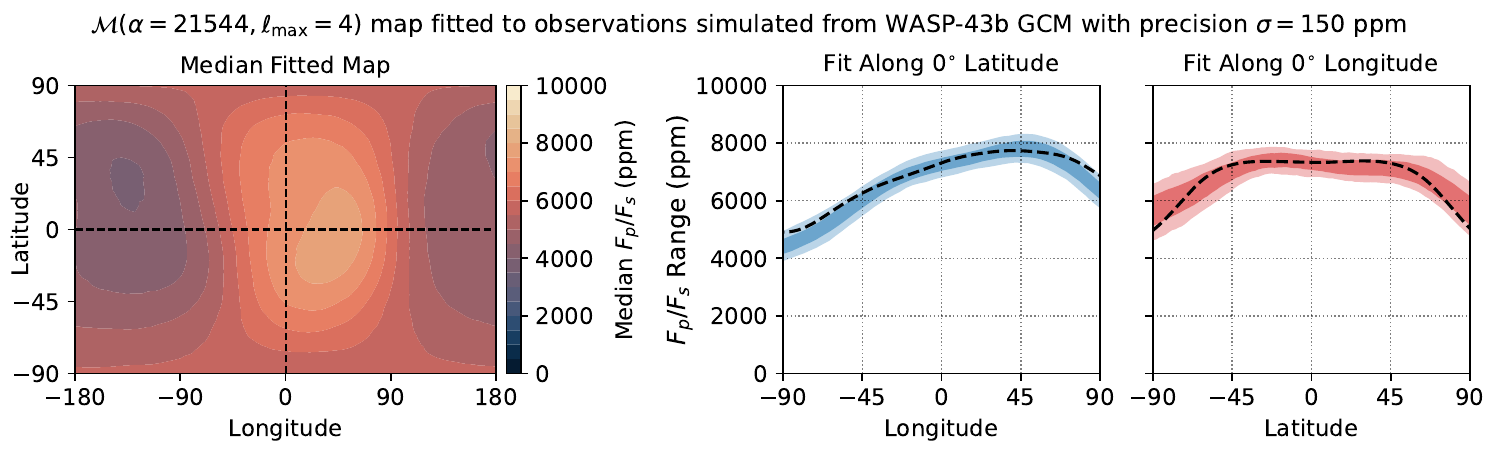}%

\includegraphics[width=\textwidth]{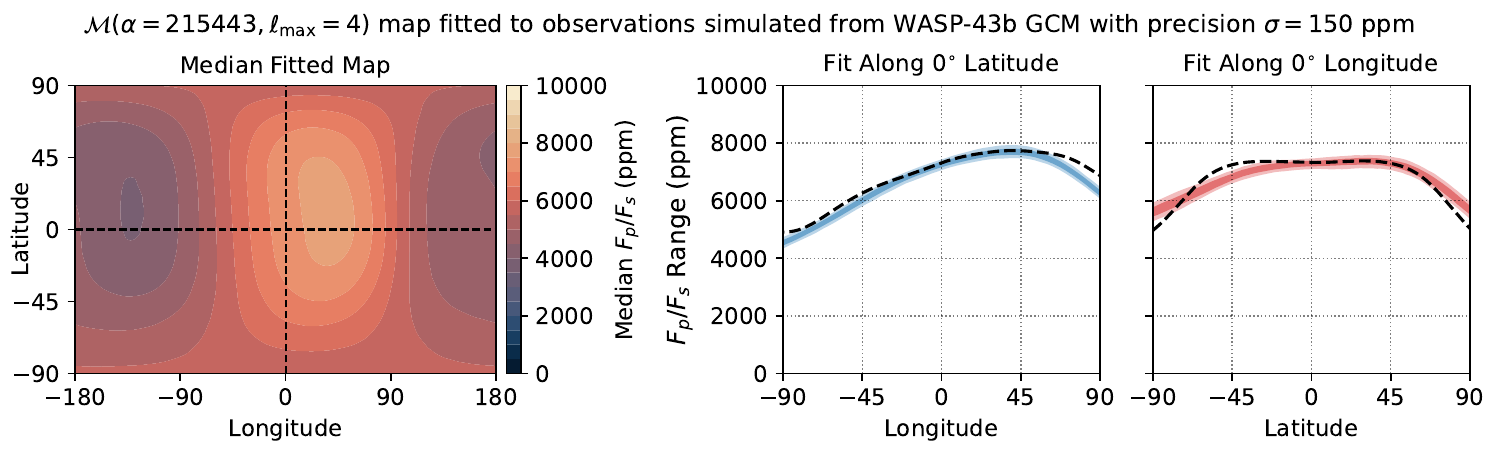}%
\caption{Three eclipse map models $\mathcal{M}(\alpha,\ell_{\mathrm{max}}=4)$ fitted to the simulated $\sigma = 150$ ppm data in Figure \ref{fig:sim_w43b_good} with variable values of $\alpha$. First row: an eclipse map model $\mathcal{M}(\alpha=2154,\ell_{\mathrm{max}}=4)$. This low value of $\alpha$ weakly penalises the likelihood of maps with a high information content (Equation \ref{eqn:L_penalty}), producing an uncertain map with a low cross-validation score in Figure \ref{fig:cv_comparison_sim_w43b}. The fitted equatorial brightness maximum is $(49^{+11}_{-36})^{\circ}$ east, compared to the true value of $40^{\circ}$ shown in Figure \ref{fig:gcm_true_l2_l4}. Second row: a model $\mathcal{M}(\alpha=21544,\ell_{\mathrm{max}}=4)$ fitted to the same data. This allows an appropriate information content in the map given the precision of the input data, giving the optimal cross-validation score in Figure \ref{fig:cv_comparison_sim_w43b}. This results in an accurate and precise measurement of the equatorial brightness maximum at $(45^{+7}_{-7})^{\circ}$ east, compared to the true value of $40^{\circ}$. Third row: a model $\mathcal{M}(\alpha=215443,\ell_{\mathrm{max}}=4)$ fitted to the same data. This high value of $\alpha$ strongly penalises the likelihood of maps with high information content, which allows too little information in the map and achieves a poor cross-validation score in Figure \ref{fig:cv_comparison_sim_w43b}. It still achieves an accurate but overly precise measurement of the position of maximum brightness on the equator, of $(38^{+7}_{-4})^{\circ}$ east, compared to the true value of $40^{\circ}$.}\label{fig:sim_gcm_l4_low_opt_high_150}
\end{figure*}

\subsection{Model comparison in simulated data}\label{sec:results_sim:model_comparison_aic_bic_cv}

We need a model comparison metric to identify the presence of an eclipse mapping signal. We define that the metric must quantitatively prefer the simplest eclipse mapping model $\mathcal{M}(\ell_{\mathrm{max}}=2)$ over the simplest Fourier series model $\mathcal{M}(n_{\mathrm{max}}=2)$. We want the metric to identify an eclipse mapping signal in the simulated datasets in Figure \ref{fig:sim_w43b_good}, but not in the simulated datasets in Figure \ref{fig:sim_w43b_bad}. Section \ref{sec:methods:aic_bic} describes the BIC, which has been used previously for eclipse mapping model comparison \citep{challener2022theresa}, and the AIC, which is an alternative metric \citep{burnham2004multimodel}. These metrics penalise the fit likelihood by a function of the number of model parameters to penalise both overfitting and underfitting.

Table \ref{tab:stats_comparison_sim_w43b} lists the difference in AIC and BIC between an eclipse map model $\mathcal{M}(\ell_{\mathrm{max}}=2)$ and a Fourier series model $\mathcal{M}(n_{\mathrm{max}}=2)$ fitted to the four simulated datasets. A positive $\Delta$AIC or  $\Delta$BIC means that the metric prefers the eclipse mapping model. This table shows that both the AIC and the BIC correctly identify an eclipse mapping signal in the 150 ppm and 250 ppm simulated data in Figure \ref{fig:sim_w43b_good}. This is because the advantage in $\chi^{2}$ for the eclipse mapping model is so high for each dataset that it outweighs the penalty for its increased number of parameters in each case.

Both the AIC and BIC also correctly identify the lack of an eclipse mapping signal in the imprecise 2000 ppm data in Figure \ref{fig:sim_w43b_bad}, which is fitted just as well by the Fourier series model. In this case, the $\chi^{2}$ value is essentially the same for the eclipse mapping model and the Fourier series model (the best value is actually slightly better for the Fourier series model, but the difference is not significant). The penalty for the increased number of parameters in the eclipse mapping model then gives it much worse BIC and AIC scores.

The picture is more complex for the 150 ppm data fitted with an incorrect value of $t_{0}$ in Figure \ref{fig:sim_w43b_bad}. The eclipse map model does fit the observed data more closely than the Fourier series model, because there are elements of the eclipse shape that it models correctly. However, it also incorrectly adjusts its latitudinal structure to compensate for the error in the eclipse timing, producing a highly incorrect eclipse map with spuriously high confidence. We suggest that a model comparison metric should not prefer the eclipse mapping model in this case, but both the BIC and the AIC strongly prefer the eclipse mapping model. This is because they cannot distinguish when a model achieves a good fit by an incorrect route. The penalty for the increased number of parameters is then not sufficient to outweigh the spuriously improved $\chi^{2}$ score. We suggest that this is a strength of the cross-validation score (see below), which identifies the fact that the eclipse map model is not fitting a robust and consistent signal.


The BIC and AIC also produce different answers for the relative weighting of models, and there is no clear answer about which metric is more appropriate \textit{a priori} as discussed in Section \ref{sec:methods:aic_bic}. Despite these issues, the AIC and BIC are useful improvements on the $\chi^{2}$ metric as they do attempt to penalise for overfitting. However, crucially for our purposes, the AIC and BIC cannot be used for the regularisation approach described in Section \ref{sec:methods:entropy}. We therefore turn to model comparison by cross-validation. Cross-validation lets us calculate the predictive power of the model, which applies a penalty for overfitting without needing to specify a number of model parameters. We calculate the k-fold cross-validation score as described in Section \ref{sec:methods:cv} for the simulated datasets in Figures \ref{fig:sim_w43b_good} and \ref{fig:sim_w43b_bad}, for the simplest eclipse mapping model $\mathcal{M}(\ell_{\mathrm{max}}=2)$ compared to the simplest Fourier series model $\mathcal{M}(n_{\mathrm{max}}=2)$. For this test, we do not penalise the model likelihood by the map entropy as described in Section \ref{sec:methods:entropy} (i.e. $\alpha=0$ here).

The left-hand panel of Figure \ref{fig:cv_comparison_sim_w43b} compares the cross-validation score difference between the eclipse mapping model and the Fourier series model for these three datasets. We normalise each relative cross-validation score by its standard error, so that a difference greater than 1 is significant. This shows that the eclipse map models achieve significantly better cross-validation scores for the datasets with precisions of $\sigma=150$ ppm and $\sigma=250$ ppm (Figure \ref{fig:sim_w43b_good}). As expected, this shows that the fitted eclipse map models are much better at predicting out-of-sample data than the Fourier series models, so they are not achieving better $\chi^{2}$ values by overfitting to noise.

Figure \ref{fig:cv_comparison_sim_w43b} hows that an eclipse map model fitted to the dataset in Figure \ref{fig:sim_w43b_bad} with an incorrect value of $t_{0}^{\prime} = t_{0} + 10\mathrm{s}$ achieves a lower cross-validation score than a Fourier series model. The standard error of the difference between the two scores shows that their scores are statistically consistent with each other, so the eclipse mapping model is no better or worse at predicting out-of-sample data than the Fourier series model. This means that there is no statistically robust eclipse mapping signal. This is an improvement on the BIC and the AIC, which preferred the eclipse mapping model in this case in Table \ref{tab:stats_comparison_sim_w43b}.

The eclipse mapping model also achieves a significantly worse cross-validation score than the Fourier series model for the dataset with precision $\sigma=2000$ ppm in the second row of Figure \ref{fig:sim_w43b_bad}. This is because the mapping model predicts a wide range of implausible solutions that can be very different to the omitted data, while the Fourier series model is restricted to a small range of solutions matching the shape of a uniform eclipse. 

We conclude that, for these simulated datasets, the cross-validation score correctly identifies if the fitted model is tracking the true eclipse mapping signal, instead of overfitting to any imprecise or inaccurate data points. The BIC and AIC perform less well at identifying the presence or absence of eclipse mapping signals and -- unlike the cross-validation score -- cannot be used with a regularisation parameter to optimise information content.

\subsection{Optimising the information content of fitted maps}\label{sec:results_sim:smooth_gcm}

Having used a simple eclipse mapping model $\mathcal{M}(\ell_{\mathrm{max}}=2)$ to establish the presence of an eclipse mapping signal in the dataset with $\sigma=$ 150 ppm in Figure \ref{fig:sim_w43b_good}, we can now fit a map with more complex structure. 
Figure \ref{fig:sim_gcm_l4_150} shows an eclipse map model $\mathcal{M}(\ell_{\mathrm{max}}=4)$ fitted to this dataset using spherical harmonics up to order $\ell_{\mathrm{max}}=4$. This high-order map is able to fit the true map in theory, but in practice it overfits a wide range of solutions, resulting in an impractically large uncertainty. The large uncertainty on the fitted position of the brightest point, at $(56^{+65}_{-58})^{\circ}$, reflects the range of different small-scale peaks fitted with the high-order harmonics, and shows the limitation of using this metric for anything but the largest-scale maps.

So, while there is a clear eclipse mapping signal, both the low-order map (Figure \ref{fig:sim_gcm_l2_150}) and the high-order map (Figure \ref{fig:sim_gcm_l4_150}) fail to match the true map. This issue is even worse for real data as we do not know the true map in advance, so do not know how many spherical harmonic orders are enough to fit accurately. This is the motivation behind the method to fit a high-order map with optimised information content described in Section \ref{sec:methods:entropy}. The right-hand panel of Figure \ref{fig:cv_comparison_sim_w43b} shows how the cross-validation score for an  eclipse map model $\mathcal{M}(\alpha,\ell_{\mathrm{max}}=4)$ varies as the regularisation parameter $\alpha$  in Equation \ref{eqn:L_penalty} is varied. For both datasets, there is a significant difference between the cross-validation scores of the lowest value of $\alpha$ (which tends towards the score for $\alpha=0$) and the optimal value of $\alpha$.

When $\alpha$ is very small, the fitted map is not strongly penalised for a high entropy (high spatial variability), so the very imprecise map in the first row of Figure \ref{fig:sim_gcm_l4_low_opt_high_150} produces the best likelihood (see Equation \ref{eqn:L_penalty}), and the data is overfitted so the cross-validation score is low. When $\alpha$ is very large, spatial variation is strongly penalised so the fit in the third row of Figure \ref{fig:sim_gcm_l4_low_opt_high_150} produces the best likelihood, and the data is underfitted so the cross-validation score is low. For the optimal value of $\alpha$, the second row of Figure \ref{fig:sim_gcm_l4_low_opt_high_150} produces the best likelihood, achieving the best cross-validation score as it has an appropriate information content. The fitted map matches the data and also allows an appropriate amount of variation as well as appropriate average gradients and feature sizes for the true map. Therefore, cross-validation gives us a data-driven method to find the optimal value of $\alpha$, and the corresponding optimal information content for the map, with no prior knowledge of the true map. We suggest that as a shortcut, choosing and applying a ``reasonable'' value of $\alpha$ is a quick way to produce a better result than a $\mathcal{M}(n_{\mathrm{max}}=4)$ model, especially if the reasonable value is derived from a previous similar observation. However, deriving the optimal value of $\alpha$ manually using cross-validation for each dataset will be best, although time-consuming.

\begin{figure*}
\centering 
  \includegraphics[width=.9\textwidth]{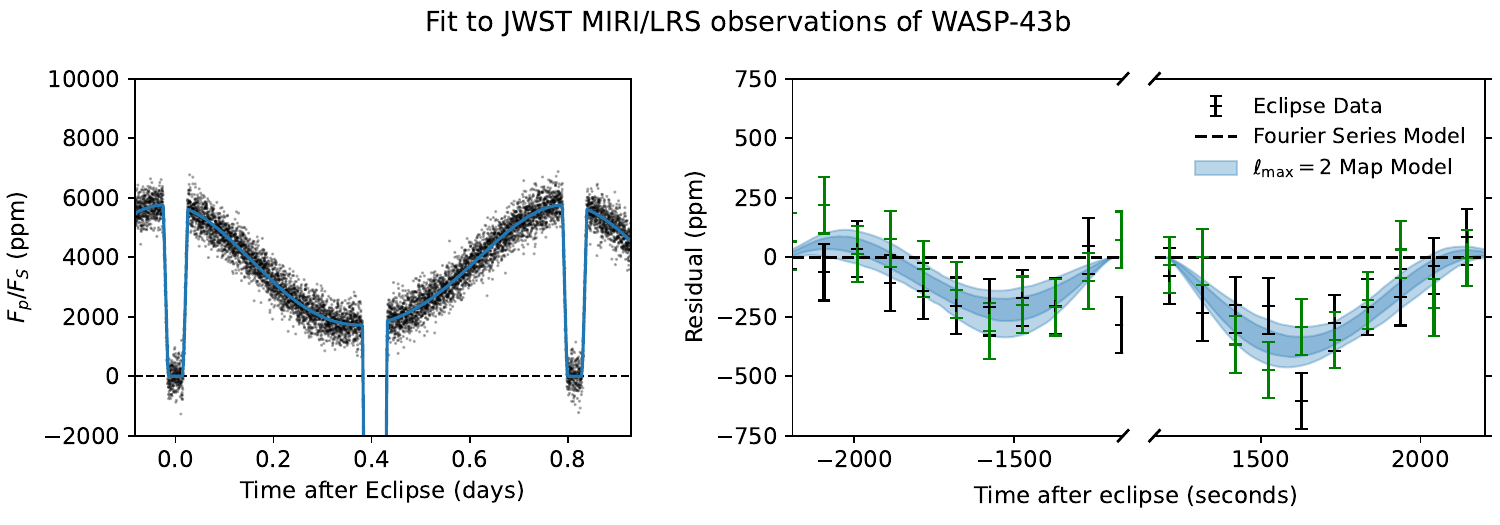}%
\caption{The thermal emission from 5 to 10.5 $\mu$m from the hot Jupiter WASP-43b observed  by JWST MIRI/LRS \citep{bell2024nightside}. The layout of panels is the same as Figure \ref{fig:sim_w43b_good}. The black points show the first eclipse and the green points show the second eclipse, binned every 10 points. The eclipse map model $\mathcal{M}(\ell_{\mathrm{max}}=2)$ fits the eclipse shape much better than the Fourier series model $\mathcal{M}(n_{\mathrm{max}}=2)$. This is confirmed by the higher cross-validation score for the eclipse map model for this dataset shown in Figure \ref{fig:obs_cv_compare} and listed in Table \ref{tab:obs_fit_stats}. The data points in the right-hand panel appear slightly different to the data points in \citet{hammond2024wasp43b} because here we bin them with a slightly longer time cadence.}\label{fig:obs_w43b_miri}
\end{figure*}
\begin{figure*}
\centering 
  \includegraphics[width=.9\textwidth]{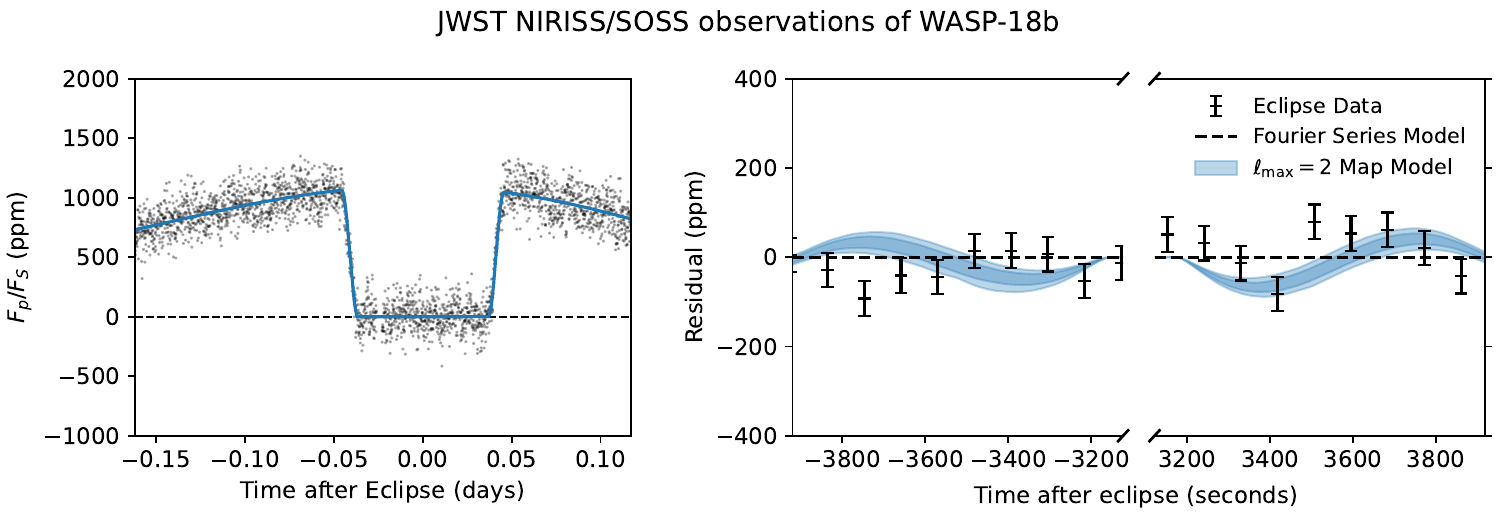}%
\caption{The thermal emission from 0.8 to 2.8 $\mu$m from the hot Jupiter WASP-18b observed by JWST NIRISS/SOSS \citep{coulombe2023broadband}. The data in the right-hand panel is binned every 10 points. There is some residual deviation in the data from the Fourier series model $\mathcal{M}(n_{\mathrm{max}}=2)$, but it is not fitted well by the eclipse mapping model $\mathcal{M}(\ell_{\mathrm{max}}=2)$. Figure \ref{fig:obs_cv_compare} suggests there is no significant eclipse mapping signal for this dataset, as the eclipse mapping model is no better than the Fourier series model at predicting out-of-sample sections of the eclipse.}\label{fig:obs_w18b_niriss}
\centering 
  \includegraphics[width=.9\textwidth]{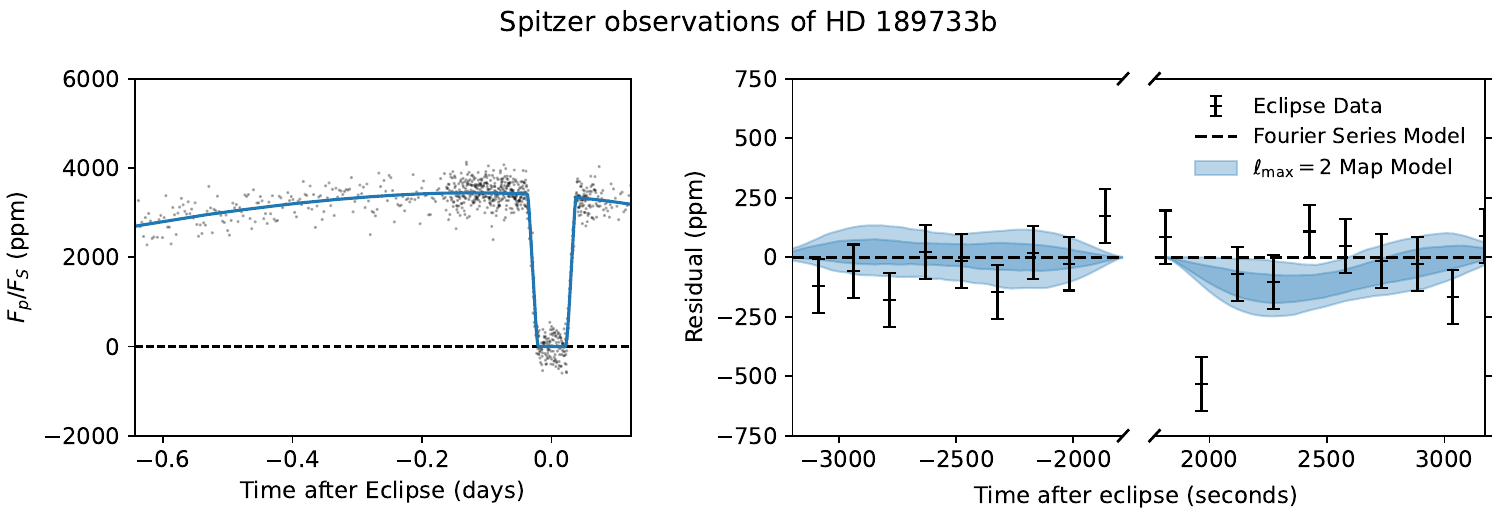}%
\caption{The observed thermal emission at 8 $\mu$m from the hot Jupiter HD 189733b observed by the Spitzer Space Telescope \citep{majeau2012two}. The data in the right-hand panel is binned every 5 points. There is some residual deviation from the Fourier series model $\mathcal{M}(n_{\mathrm{max}}=2)$, but it is not fitted well by the eclipse mapping model $\mathcal{M}(\ell_{\mathrm{max}}=2)$. Figure \ref{fig:obs_cv_compare} suggests there is no significant eclipse mapping signal for this dataset, as the eclipse mapping model is no better than the Fourier series model at predicting out-of-sample sections of the eclipse.}\label{fig:obs_hd189_spitzer}
\end{figure*}

\begin{table*}
    \centering
    \begin{tabular}{lcccc}
         \textbf{Dataset} & $\Delta \chi^{2}$ &  $\Delta$BIC & $\Delta$AIC &$\Delta$CV  \\
         \hline
         WASP-43b (JWST MIRI/LRS) & +103.4 & +58.2 & +93.4 &  $+0.060 \pm 0.008$   \\
         WASP-18b (JWST NIRISS SOSS) & -18.2 & -57.69 & -28.2 & $-0.054 \pm 0.006$  \\
         HD 189733b (Spitzer) & +0.92 & -32.98  & -9.08 &  $ -0.067 \pm 0.008$ \\
    \end{tabular}
    \caption{Statistical metrics comparing eclipse mapping models $\mathcal{M}(\ell_{\mathrm{max}}=2)$ to Fourier series models $\mathcal{M}(n_{\mathrm{max}}=2)$, for the observational datasets in Figures \ref{fig:obs_w43b_miri}, \ref{fig:obs_w18b_niriss}, and \ref{fig:obs_hd189_spitzer}. We propose that $\Delta$CV is the best metric for identifying an eclipse mapping signal, as discussed in Section \ref{sec:results_sim}. Only the JWST MIRI/LRS WASP-43b dataset in Figure \ref{fig:obs_w43b_miri} has statistical evidence for an eclipse mapping signal by this metric. Note that the BIC and AIC values for the WASP-43b dataset are slightly different to the equivalent models fitted in \citet{hammond2024wasp43b} because here we do not discard the data in the transit, and do not include an additional parameter to fit the baseline of the eclipses. The WASP-18b dataset prefers the Fourier series model (the null hypothesis, with no mapping information) by all metrics. The HD 189733b dataset prefers the Fourier series model by all metrics except $\Delta \chi^{2}$, but we suggest this is due to overfitting.}
    \label{tab:obs_fit_stats}
\end{table*}

\begin{figure*}
\centering
  \includegraphics[width=\textwidth]{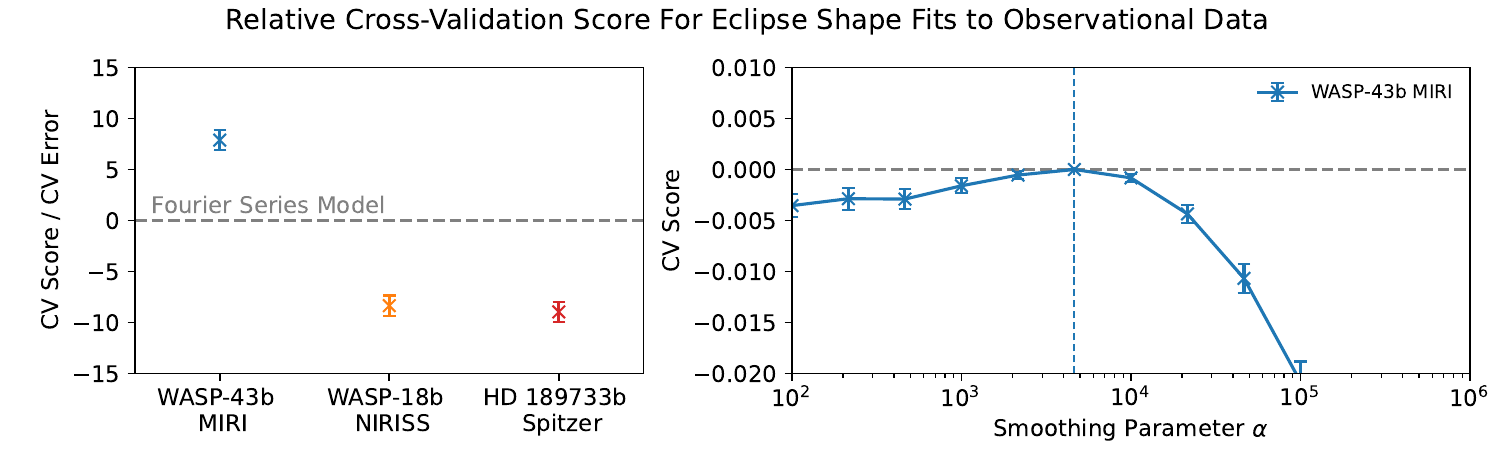}%
\caption{Left panel: the difference in cross-validation score (Equation \ref{eqn:delta_cv}) for each dataset for the eclipse map model $\mathcal{M}(\ell_{\mathrm{max}}=2)$ compared to the Fourier series model $\mathcal{M}(n_{\mathrm{max}}=2)$. Each score is normalised by its standard error (Equation \ref{eqn:se_delta_cv}). Of the three observational datasets, only the JWST MIRI/LRS observations of WASP-43b in Figure \ref{fig:obs_w43b_miri} achieves a better cross-validation score with the eclipse map model. Right panel: the cross-validation score of the eclipse map model $\mathcal{M}(\alpha,\ell_{\mathrm{max}}=4)$ fitted to the JWST MIRI/LRS observations of WASP-43b in Figure \ref{fig:obs_w43b_miri}, with its likelihood penalised by the entropy scaled by $\alpha$ as described in Section \ref{sec:methods:entropy}. The best cross-validation score is achieved by setting $\alpha=4641.6$, which produces the map in Figure \ref{fig:obs_w43b_miri_opt}.}\label{fig:obs_cv_compare}
\end{figure*}

\section{Observational Results}\label{sec:results_obs}

In this section, we take the method demonstrated on simulated data in Section \ref{sec:results_sim}, and apply it to three observational datasets. These datasets are the JWST MIRI/LRS observations of WASP-43b in Figure \ref{fig:obs_w43b_miri} \citep{bell2024nightside}, the JWST NIRISS/SOSS observations of WASP-18b in Figure \ref{fig:obs_w18b_niriss} \citep{coulombe2023broadband}, and the Spitzer Space Telescope observations of HD 189733b in Figure \ref{fig:obs_hd189_spitzer} \citep{majeau2012two}. The left-hand panel of these plots shows the raw observations, fitted by an eclipse map model $\mathcal{M}(\ell_{\mathrm{max}}=2)$. The right-hand panel of these plots shows the residual signal in ingress and egress.

\subsection{Identifying eclipse mapping signals in observed data}\label{sec:obs:identify_signals}

\begin{figure*}
\centering 
  \includegraphics[width=\textwidth]{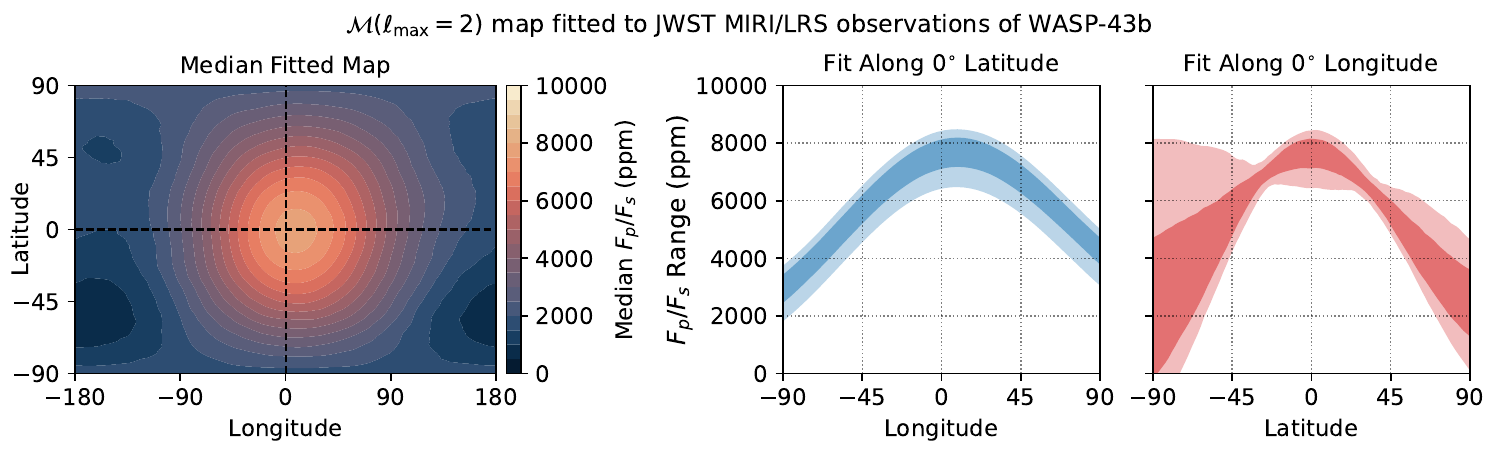}%
\caption{An eclipse map fitted to the data in Figure \ref{fig:obs_w43b_miri}, using a model $\mathcal{M}(\ell_{\mathrm{max}}=2)$. Table \ref{tab:obs_fit_stats} and Figure \ref{fig:obs_cv_compare} show how this model is strongly preferred over the Fourier series model (the null hypothesis) by the cross-validation metric. \citet{hammond2024wasp43b} discussed how the structure of this $\ell_{\mathrm{max}}=2$ map is limited by its limited spatial freedom.}\label{fig:obs_w43b_miri_l2}
\end{figure*}
\begin{figure*}
\centering 
  \includegraphics[width=\textwidth]{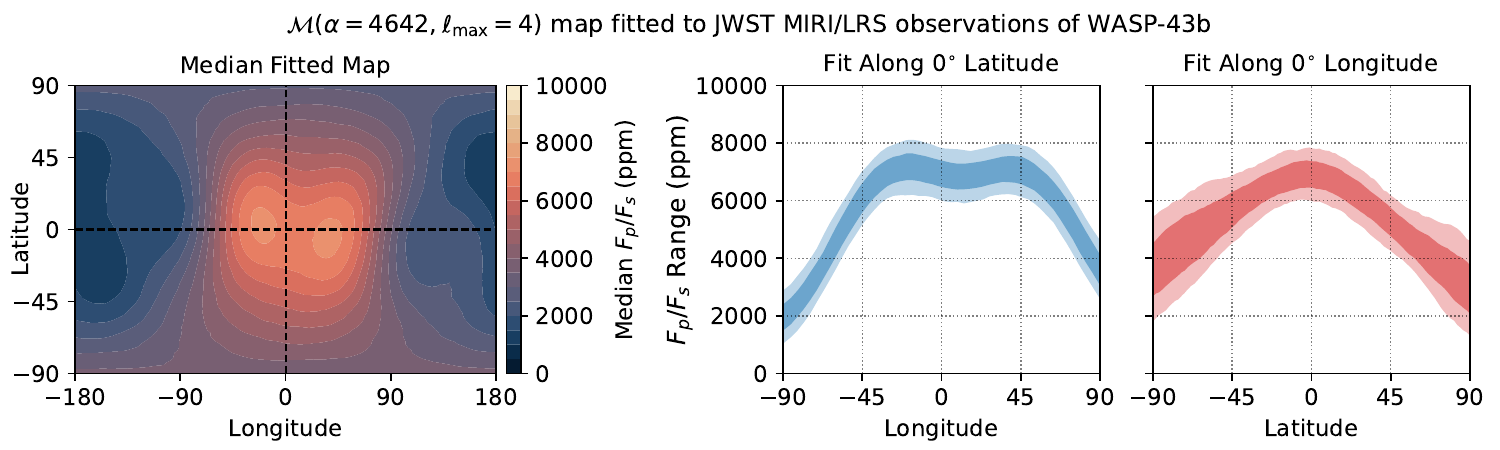}%
\caption{An eclipse map fitted to the data in Figure \ref{fig:obs_w43b_miri} using a model $\mathcal{M}(\alpha=4641.6,\ell_{\mathrm{max}}=4)$. This value of $\alpha$ gives the optimal cross-validation score in Figure \ref{fig:obs_cv_compare}. We suggest that this value imposes an information content which reflects a combination of the spatial scale of the true map and the level of mapping precision achievable given the precision of the data itself.}\label{fig:obs_w43b_miri_opt}
\end{figure*}

We apply our statistical method defined in Section \ref{sec:methods} to determine if there is quantitative evidence for an eclipse map in each dataset. The left-hand panel of Figure \ref{fig:obs_cv_compare} shows the difference in cross-validation score for an eclipse map model $\mathcal{M}(\ell_{\mathrm{max}}=2)$ compared to a Fourier series model $\mathcal{M}(n_{\mathrm{max}}=2)$, for each of the observational datasets in Figures \ref{fig:obs_w43b_miri}, \ref{fig:obs_w18b_niriss}, and \ref{fig:obs_hd189_spitzer}. We do not apply a penalty to the likelihood of a fitted map based on its entropy for these $\mathcal{M}(\ell_{\mathrm{max}}=2)$ models; we only later consider a non-zero $\alpha$ once we have established the presence of a mapping signal with this initial test. The WASP-43b dataset has a positive $\Delta$CV value, showing significant evidence for an eclipse mapping signal, matching the clear fit to the large residual in the data in Figure \ref{fig:obs_w43b_miri}. Figure \ref{fig:obs_w43b_miri_l2} shows an eclipse map model $\mathcal{M}(\ell_{\mathrm{max}}=2)$ fitted to this dataset. The posterior distribution of this fitted map is slightly different to the $\ell_{\mathrm{max}}=2$ map in \citet{hammond2024wasp43b} because the prior placed on the brightness of each pixel is wider in Figure \ref{fig:obs_w43b_miri_l2}. 

On the other hand, the WASP-18b dataset has a negative $\Delta$CV value in Figure \ref{fig:obs_cv_compare}, suggesting that there is no statistical evidence for an eclipse mapping signal by this metric. The WASP-18b dataset has a comparable signal-to-noise ratio to the WASP-43b dataset, so we would expect that it would be possible to resolve a mapping signal. The mismatch between the fitted and observed residual signal in Figure \ref{fig:obs_w18b_niriss} implies that the longitudinal structure imposed by the out-of-eclipse phase curve is inconsistent with the structure implied by the eclipse shape. This may indicate an unresolved systematic error in the astrophysical model or the instrumental model, like the inaccurate simulated data in the top row of Figure \ref{fig:sim_w43b_bad}. It is also possible that the day-side has a complex structure that produces a smoothly varying phase curve but a largely flat structure that requires higher-order harmonics than $\ell_{\mathrm{max}}=2$ to resolve. The eclipse mapping model $\mathcal{M}(\ell_{\mathrm{max}}=2)$ (shown in Supplementary Figure \ref{fig:s5_w18b}) actually achieves a worse $\chi^{2}$ score than the Fourier series model $\mathcal{M}(n_{\mathrm{max}}=2)$ in Table \ref{tab:obs_fit_stats} for this dataset.


The HD 189733b dataset also has a negative $\Delta$CV values in Figure \ref{fig:obs_cv_compare}, again suggesting that there is no statistical evidence for an eclipse mapping signal by this metric. This conclusion matches the visual evidence in Figures \ref{fig:obs_w18b_niriss} and \ref{fig:obs_hd189_spitzer}, where there is no clear fit to a residual signal. We suggest that the precision of the HD 189733b dataset may be too low to identify an eclipse mapping signal, like the imprecise simulated dataset in Figure \ref{fig:sim_w43b_bad}. Table \ref{tab:obs_fit_stats} shows that the eclipse mapping model $\mathcal{M}(\ell_{\mathrm{max}}=2)$ (shown in  Supplementary Figure \ref{fig:s6_hd189}) achieves a slightly better $\chi^{2}$ score for the HD 189733b data than the Fourier series model $\mathcal{M}(n_{\mathrm{max}}=2)$. We suggest this is due to overfitting to noise in the eclipse shape, rather than matching an underlying signal. This overfitting is penalised by the cross-validation score, which prefers the simpler Fourier series model. This suggests that the two-dimensional structure implied by this map is due to a combination of the out-of-eclipse phase curve providing longitudinal structure and limits on the latitudinal structure due to the priors (including the need for positive brightness) placed on the sampled parameters. 



The signs of the relative cross-validation scores in Table \ref{tab:obs_fit_stats} match the signs of the BIC and AIC scores and are consistent with the presence or absence of visual signals in Figures \ref{fig:obs_w43b_miri}, \ref{fig:obs_w18b_niriss}, and \ref{fig:obs_hd189_spitzer}. We conclude that the $\Delta$CV score provides strong statistical evidence for an eclipse mapping signal in the WASP-43b dataset only. We now proceed to the second part of our method for this dataset, fitting an $\mathcal{M}(\alpha,\ell_{\mathrm{max}}=4)$  model with its information content optimised by the method described in Section \ref{sec:methods:entropy}.

\subsection{An optimised map fitted to observations of WASP-43b}\label{sec:obs:w43b_map}

As discussed in Section \ref{sec:results_sim}, spherical harmonics up to $\ell_{\mathrm{max}}=2$ have limited spatial freedom and may not be able to fit the true map for the observed WASP-43b dataset. We therefore fit a new model $\mathcal{M}(\alpha,\ell_{\mathrm{max}}=4)$ with spherical harmonics up to $\ell_{\mathrm{max}}=4$, with its information content determined by a regularisation parameter $\alpha$ (Equation \ref{eqn:L_penalty}). The right-hand panel of Figure \ref{fig:obs_cv_compare} shows how the cross-validation score for this eclipse map varies as a function of $\alpha$. There is a peak at $\alpha=4641.6$, corresponding to the optimal amount of map information that allows an accurate fit to the data without overfitting to the noise. 

Figure \ref{fig:obs_w43b_miri_opt} shows the resulting $\ell_{\mathrm{max}}=4$ map with $\alpha=4641.6$. The map has a flatter emission structure near the substellar point in longitude than the $\ell_{\mathrm{max}}=2$ map. Its location of maximum brightness is at $(15^{+27}_{-38}) ^{\circ}$ longitude and $(-3^{+13}_{-13}) ^{\circ}$ latitude, relative to the substellar point. There appear to be two peaks in the dayside brightness distribution in the median map in Figure \ref{fig:obs_w43b_miri_opt}, but we suggest that the posterior distribution shows this is not a significant feature. The large uncertainty of the longitudinal position of the brightness maxima reflects the flat emission structure along the equator near the substellar point, rather than a range of distinct peaks, so the assumed Gaussian distribution that leads to the large uncertainty may not be a good assumption here. The latitudinal structure contributes significantly to the eclipse shape, as identified in \citet{hammond2024wasp43b}, and is consistent with a brightness maximum on the equator. There is a small asymmetry in the latitudinal structure which is not statistically significant, given the width of the posterior distribution. The latitudinal structure varies more smoothly than the longitudinal structure, which transitions from a low gradient near the substellar point to sharp gradients near the terminators.

\citet{hammond2024wasp43b} suggested that a uniform longitudinal structure near the substellar point could be caused by dayside stationary Rossby waves. These waves appear in some of the GCM simulations in \citet{hammond2024wasp43b} and produce a less sharply peaked longitudinal structure \citep{lewis2022temperature}. However, they also produce a two-peaked latitudinal structure with maxima north and south of the equator \citep{matsuno1966quasi}, but the data are not sufficiently precise to resolve or rule out a two-peaked structure in Figure \ref{fig:obs_w43b_miri_opt}. More observations might constrain the dayside structure strongly enough to detect such structures. We note that some of the the ``observable'' GCM maps in \citet{hammond2024wasp43b} have their two-peaked structures and their chevron-shaped structures suppressed compared to the original GCM simulation, making them more similar to the optimised map we derive here.

\subsection{Comparison to theoretical expectations}

To put these results in context, we compare them to the precision predicted by the analytic ``Eclipse Mapping Metric'' in \citet{boone2024analytical}. To calculate the predicted $\mathrm{EMM}$ for the most precise simulated WASP-43b data in Figure \ref{fig:sim_w43b_good}, we set the error per point to $\sigma=150$ ppm (note that \citet{boone2024analytical} denotes the error per point as $\sigma_{1}$), the number of points to $N_{\mathrm{points}}=7000$, the peak flux as $F_{0}=7000$ ppm, the inclination  to $i=82.106^{\circ}$, the semi-major axis to $a=4.859 R_{*}$, and the planetary radius to $R_{p}=0.15839R_{*}$. This results in an eclipse mapping metric of $\mathrm{EMM}=23^{\circ}$, which predicts the smallest scale resolvable by an eclipse map. This is consistent with Figure \ref{fig:sim_gcm_l4_low_opt_high_150}, where the map can resolve variations in latitude and longitude on the scale of tens of degrees.

We use the same set of parameters to calculate the $\mathrm{EMM}$ for the real WASP-43b data in Figure \ref{fig:obs_w43b_miri}, except for the precision per point which we change to be $\sigma=250$ ppm, representing the accuracy achieved by averaging two eclipses (the precision of a single point is $\sigma\approx370$ ppm \citep{bell2024nightside}). This gives $\mathrm{EMM}=27^{\circ}$, implying that the map in Figure \ref{fig:obs_w43b_miri_opt} should also be able to resolve variations on the scale of tens of degrees, although with less precision than the best-case simulated map. To achieve the same eclipse mapping metric, we would need to observe five such eclipses of WASP-43b.

To estimate the $\mathrm{EMM}$ for the WASP-18b dataset in Figure \ref{fig:obs_w18b_niriss},  we set the error per point to $\sigma=121$ ppm, the number of points for a full phase curve to $N_{\mathrm{points}}=9184$, the peak flux as $F_{0}=1000$ ppm, the inclination to $i=84.39^{\circ}$, the semi-major axis to $a=3.483 R_{*}$, and the planetary radius to $R_{p}=0.0943R_{*}$. This results in $\mathrm{EMM}=62^{\circ}$, which is significantly hampered by the low impact parameter of the eclipse. Note that this is a less precise $\mathrm{EMM}$ than the predicted value of $\mathrm{EMM}=47^{\circ}$ for WASP-18b in \citet{boone2024analytical}, which used an estimated value for the precision of NIRISS observations of WASP-18b that was more precise than the actual observations. The eclipse mapping metric for the longitudinal direction on WASP-18b is $\mathrm{EMM}_{x}=21^{\circ}$, predicting that the effect of the longitudinal structure on the eclipse shape should be resolvable in this case. The mismatch between the longitudinal structure implied by the phase curve, and the observed eclipse shape, implies that there may be unresolved systematic errors affecting the eclipse shape, or that the dayside may have a structure that is very uniform and is poorly represented by $\ell_{\mathrm{max}}=2$ harmonics. Observing a full phase curve including a transit would help in fitting astrophysical and instrumental models for an observation of WASP-18b, as separating time-correlated noise from physical phase variations is difficult without a full phase curve constrained by periodicity.

We also estimate the $\mathrm{EMM}$ for the HD 189733b data in Figure \ref{fig:obs_hd189_spitzer}. We set the error per point to $\sigma=252$ ppm, the number of points for a full phase curve to $N_{\mathrm{points}}=7160$ (calculated from the higher-cadence data which covers the eclipse), the peak flux to $F_{0}=3500$ppm, the inclination to $i=85.68^{\circ}$, the semi-major axis to $a=8.863 R_{*}$, and the planetary radius to $R_{p}=0.153R_{*}$. This results in $\mathrm{EMM}=36^{\circ}$, which is lower than the WASP-43b dataset but still implies that mapping may be possible. The lack of a discernible residual shape in Figure \ref{fig:obs_hd189_spitzer} and the related lack of a detection of a mapping signal via cross-validation implies that the precision of the data is too low or there are unresolved systematic errors.

Based on these results, we suggest that the eclipse mapping metric defined by \citet{boone2024analytical} should be at least $\mathrm{EMM}=30^{\circ}$ to resolve an eclipse map. Observing three additional eclipses of WASP-43b would make its potential resolution comparable to our highest-precision simulated data.

\section{Conclusions}

In this study, we have presented a new method for fitting eclipse maps to observations of exoplanets. This method provides two improvements on previous work. Firstly, it provides a data-driven method to assess when an eclipse mapping signal is resolvable in a dataset. Secondly, it provides a data-driven method to assess the appropriate amount of spatial information to include when fitting a map to this signal. The method has two stages:

\begin{enumerate}
    \item Test for the presence of an eclipse mapping signal by fitting a low-order Fourier series model $\mathcal{M}(n_{\mathrm{max}}=2)$ and a low-order eclipse map model $\mathcal{M}(\ell_{\mathrm{max}}=2)$, and comparing their cross-validation scores. There is an eclipse mapping signal if the eclipse map model has a better cross-validation score. The residual signal in ingress and egress should confirm this. We suggest an $\ell_{\mathrm{max}}=2$ eclipse map model and an $n_{\mathrm{max}}=2$ Fourier series model because these are simple enough to fit alongside instrumental systematics and have the same freedom to fit the light curve outside the eclipse.
    \item Once an eclipse mapping signal is identified, re-fit the map with as much spatial freedom as possible and prevent overfitting by penalising the likelihood by a factor proportional to the entropy of the map. The scale of this factor is tuned to optimise the cross-validation score of the fitted map, which identifies the correct penalty size that preserves an appropriate amount of information content given the precision of the data. An eclipse map model $\mathcal{M}(\alpha,\ell_{\mathrm{max}}=4)$, oversampled in pixel space by a factor of 3 \citep{luger2019starry}, balances sampling time and spatial freedom.
\end{enumerate}

We used k-fold cross-validation to calculate the cross-validation score, using folds with a duration of one-fifth of the eclipse ingress or egress, to test the ability of the fitted maps to predict an appropriate amount of spatial information. This metric penalised both underfitting with too few degrees of freedom and overfitting with too many degrees of freedom. To measure this score in a reasonable time, we restricted the sampled folds to cover the eclipse ingress and egress as well as an equally long section of the out-of-eclipse phase curve on either side of the eclipse.

We conclude that this method is an improvement on fitting eclipse maps using spherical harmonics alone without entropy-based regularisation. Our approach is, we hope, a complementary method to the widely used eigenmapping fitting method \citep{rauscher2018more}. The eigenmapping method adjusts the number of functions used to fit the lightcurve to optimise a statistical metric, while our new method adjust the information content of the map to optimise a different statistical metric. We showed that our new method correctly identified the presence of an eclipse mapping signal in accurate and precise simulated datasets and the absence of an eclipse mapping signal in inaccurate or imprecise simulated datasets. We then fitted a map of appropriate precision to the datasets containing eclipse mapping signals.

We applied our method to three observational datasets -- WASP-43b, WASP-18b, and HD 189733b. The cross-validation test suggested that there was only a robust eclipse mapping signal in the WASP-43b dataset. We used this to derive a new eclipse map of WASP-43b, which makes use of the increased degrees of freedom of a higher-order eclipse map model $\mathcal{M}(\alpha,\ell_{\mathrm{max}}=4)$ to fit a flatter brightness structure near the substellar point compared to previous low-order maps \citep{hammond2024wasp43b}. This flat brightness structure could be consistent with stationary wave structures seen in previous GCM simulations. However, the lack of a double-peaked latitudinal structure is inconsistent with the structure of these waves in the simulations. We conclude that there is evidence for a small eastward dynamical hot-spot shift, a longitudinally broad and uniform hot-spot, and a smoothly varying latitude structure with a robustly detected equator-pole brightness difference. 

The lower cross-validation scores for the eclipse mapping models for the WASP-18b and HD 189733b datasets suggest that there is no eclipse mapping signal in these datasets. Specifically, they show that the mapping models have lower predictive ability for left-out data than the Fourier series model. We suggest that the HD 189733b dataset has too low precision to resolve a mapping signal. The WASP-18b dataset may have unresolved systematic effects that prevent an accurate map being fitted to its sufficiently high-precision data, or may have a complex day-side structure with largely uniform regions that are not represented well by low-order spherical harmonics.

We conclude that eclipse mapping requires data of comparable quality to the WASP-43b dataset and that care should be taken when fitting maps to future datasets. Future work could improve the sampling method to allow fitting with more pixels and more degrees of spatial freedom. An alternative model comparison metric would be the Bayesian Evidence as estimated by nested sampling algorithms, which would be prohibitively slow to calculate with our method but could be used in future work. The k-fold cross-validation method could also be varied to see if there is an alternative cross-validation metric that is even better suited to comparing models of eclipse shape.

\section*{Acknowledgements}

We thank Louis-Philippe Coulombe, Adina Feinstein, and Nestor Espinoza for providing access and insight to the WASP-18b NIRISS dataset. We thank Eric Agol and Nicolas Cowan for providing access and insight to the HD 189733b Spitzer dataset. M.H. acknowledges support from Christ Church, University of Oxford. T.J.B.~acknowledges funding support from the NASA Next Generation Space Telescope Flight Investigations program (now JWST) via WBS 411672.07.05.05.03.02. Some of the data presented in this article were obtained from the Mikulski Archive for Space Telescopes (MAST) at the Space Telescope Science Institute, which is operated by the Association of Universities for Research in Astronomy, Inc., under NASA contract NAS 5-03127.

\section*{Data Availability}

The simulated and observational datasets, as well as the code used to generate, fit, and plot them, are available at https://zenodo.org/records/11394012.



\bibliographystyle{mnras}
\bibliography{main} 



\begin{figure*}
\centering 
  \includegraphics[width=\textwidth]{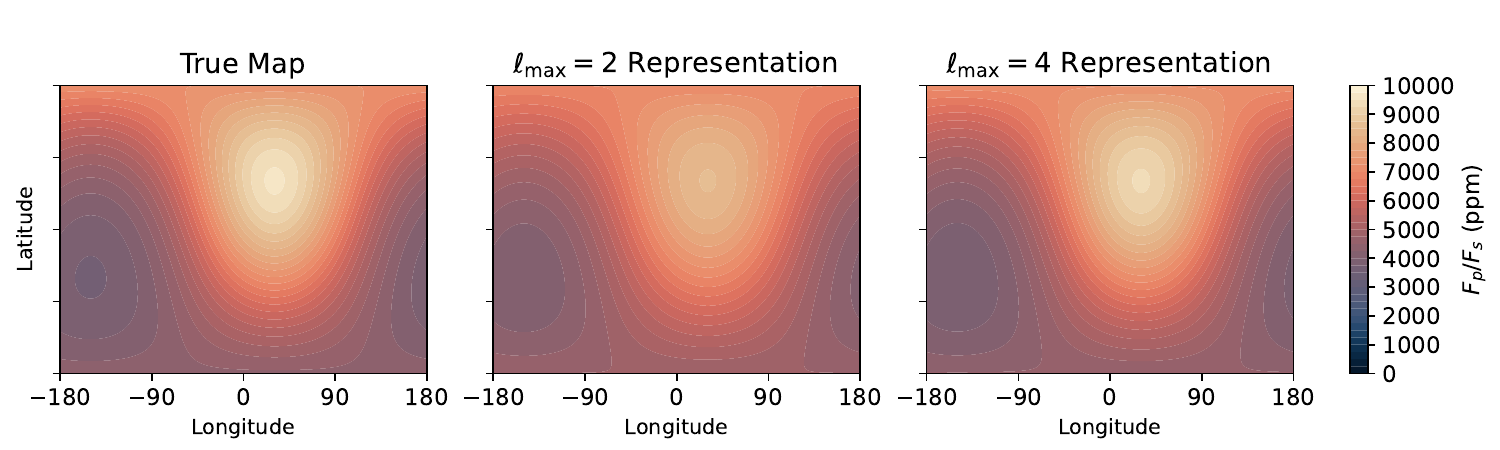}%
\caption{First panel: the artificial emission map defined by Equation \ref{eqn:perturbation_map}. Second panel: the map represented by $\ell_{\mathrm{max}}=2$ spherical harmonics, showing how this produces a slightly wider perturbation. Third panel: the map represented by $\ell_{\mathrm{max}}=4$ spherical harmonics, showing how they can capture most of its true spatial structure.}\label{fig:bump_true_l2_l4}
\end{figure*}

\appendix

\section{Additional demonstration of optimising mapping information content}\label{sec:results_sim:smooth_test}

Section \ref{sec:results_sim:smooth_gcm} showed how to fit a map with optimised information content to a simulated dataset from a numerical model of WASP-43b. In this appendix, we demonstrate the model for a different input map, showing how it selects a different spatial scale for the fitted map. We define an artificial flux map $F_{p}/F_{S}$ that is entirely uniform except for a Gaussian perturbation:
\begin{equation}\label{eqn:perturbation_map}
    F_{p}/F_{S} = F_{0} (1 + 2 e^{(-(r/r_{0})^{2}))},
\end{equation}
where $F_{0}=3000$ ppm is the magnitude of the map, $r$ is the angular distance along the surface away from a point which we choose to be $30^{\circ}$ east and $30^{\circ}$ north of the substellar point, and $r_{0}=60^{\circ}$ is the scale of the perturbation. Figure \ref{fig:bump_true_l2_l4} shows this map, as well as $\ell_{\mathrm{max}}=2$ and $\ell_{\mathrm{max}}=4$ representations of it. The $\ell_{\mathrm{max}}=2$ representation can capture the approximate shape of the true map, but fails to capture the magnitude of the perturbation correctly, while the $\ell_{\mathrm{max}}=4$ representation better matches the magnitude.

\begin{figure*}
\centering 
  \includegraphics[width=\textwidth]{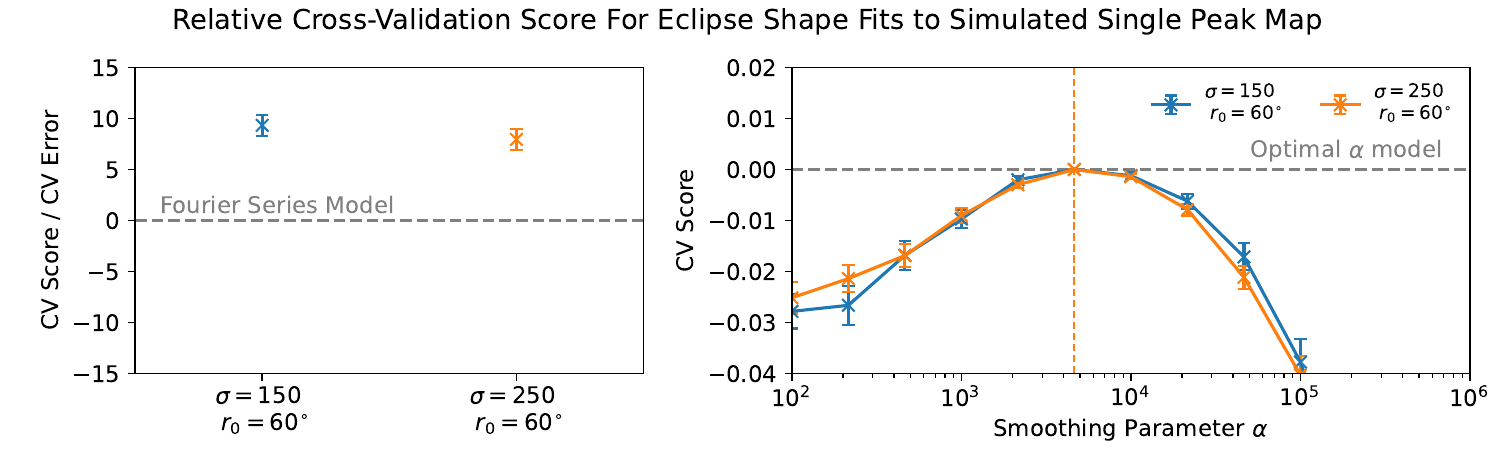}%
\caption{Left panel: The cross-validation scores (Equation \ref{eqn:delta_cv}) for eclipse map models $\mathcal{M}(\ell_{\mathrm{max}}=2)$ fitted to the simulated datasets defined by Equation \ref{eqn:perturbation_map}, relative to Fourier series models $\mathcal{M}(n_{\mathrm{max}}=2)$. Both datasets have strong eclipse mapping signals, identified by the better cross-validation scores for the eclipse mapping models $\mathcal{M}(\ell_{\mathrm{max}}=2)$. The relative scores are normalised by their standard errors (Equation \ref{eqn:se_delta_cv}). Right panel: The relative cross-validation scores for eclipse map models $\mathcal{M}(\ell_{\alpha,\mathrm{max}}=4)$, as a function of $\alpha$ (Equation \ref{eqn:L_penalty}). The optimal value of $\alpha$ for both maps is smaller than in Figure \ref{fig:cv_comparison_sim_w43b}.}\label{fig:sim_blob_cv_compare}
\end{figure*}

We generate simulated light curves following the methodology in Section \ref{sec:methods:simulation}, assuming the system parameters of WASP-43b again, and applying Gaussian noise with standard deviations of 150 ppm and 250 ppm. Figure \ref{fig:sim_blob_cv_compare} shows our mapping method applied to the two resulting datasets. The left-hand panel shows their cross-validation scores as defined in Section \ref{sec:methods:cv} for eclipse map models $\mathcal{M}(\ell_{\mathrm{max}}=2)$, relative to the scores for Fourier series models $\mathcal{M}(n_{\mathrm{max}}=2)$. The eclipse mapping models have significantly better cross-validation scores, confirming the strong eclipse mapping signal in both datasets.

\begin{figure*}
\centering 
\includegraphics[width=\textwidth]{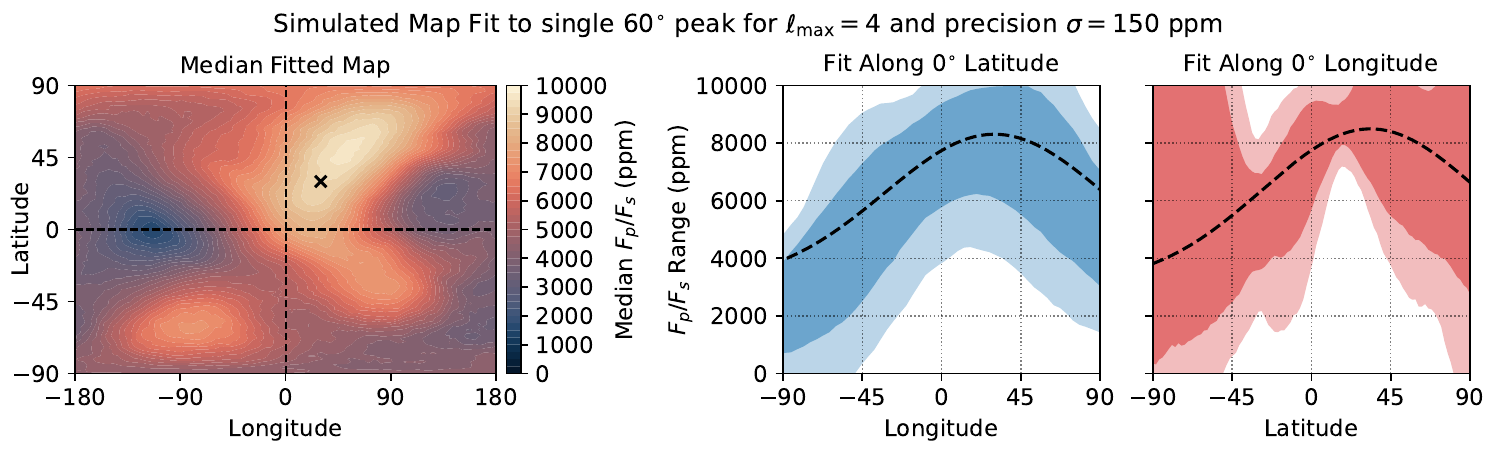}%
  
\includegraphics[width=\textwidth]{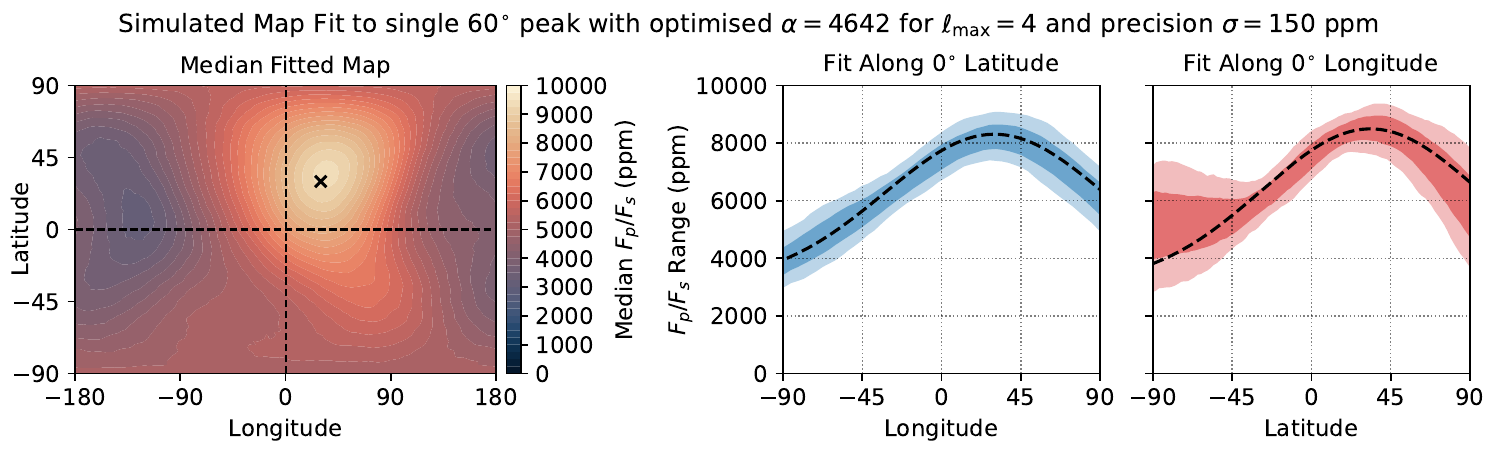}%
\caption{First row: an eclipse map model $\mathcal{M}(\ell_{\mathrm{max}}=4)$ fitted to a light curve simulated from Equation \ref{eqn:perturbation_map} with precision 150 ppm as described in Appendix \ref{sec:results_sim:smooth_test}. The crosses on the maps show the position of the peak of the Gaussian perturbation defined in Equation \ref{eqn:perturbation_map}. Fitting the the spherical harmonics alone results in a highly imprecise map. Second row: an eclipse map model $\mathcal{M}(\alpha=4642,\ell_{\mathrm{max}}=4)$ with the parameter $\alpha$ optimised according to the cross-validation score shown in Figure \ref{fig:sim_blob_cv_compare}.}\label{fig:s2_peak_150_l4_opt}
\end{figure*}

The right-hand panel of Figure \ref{fig:sim_blob_cv_compare} shows the cross-validation score for $\mathcal{M}(\ell_{\mathrm{max}}=2)$ eclipse map models fitted as described in Section \ref{sec:methods:entropy}, as a function of the regularisation parameter $\alpha$ described in Equation \ref{eqn:L_penalty}. The optimal value of $\alpha=4641.6$ is the same for both datasets, and smaller than the optimal values in Figure \ref{fig:cv_comparison_sim_w43b} for the data simulated from the GCM. The optimal value of $\alpha$ depends non-trivially on the spatial scale and precision of the dataset. We found that simulating these datasets again with different randomly generated Gaussian noise could result in different optimal values of $\alpha$, as the mapping residuals have relatively short durations and low signal to noise ratios. The optimised map in the second row of Figure \ref{fig:s2_peak_150_l4_opt} is a much more precise fit to the true map than the map in the first row which was fitted without optimising the information content.

\begin{figure*}
\centering 
  \includegraphics[width=\textwidth]{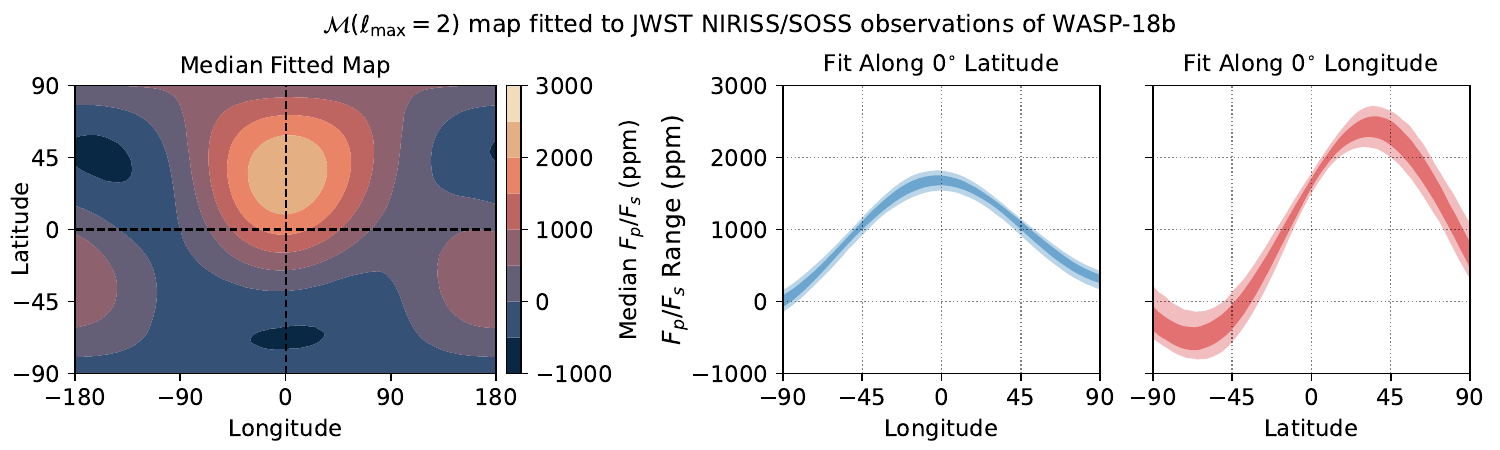}%
\caption{Supplementary Figure: an eclipse map model $\mathcal{M}(\ell_{\mathrm{max}}=2)$ fitted to the WASP-18b JWST NIRISS/SOSS observations. Section \ref{sec:obs:identify_signals} and Figure \ref{fig:obs_cv_compare} suggest that this model is not preferred over a simpler Fourier series model that does not fit an eclipse map. This model appears to produce a well-constrained posterior distribution, but this is not consistent with the eclipse shape in Figure \ref{fig:obs_w18b_niriss}. This shows why statistical metrics like cross-validation are important, as it shows that this apparently precise map model cannot predict out-of-sample sections any better than a model assuming the eclipse shape of a uniform planet.}\label{fig:s5_w18b}
\end{figure*}

\begin{figure*}
\centering 
  \includegraphics[width=\textwidth]{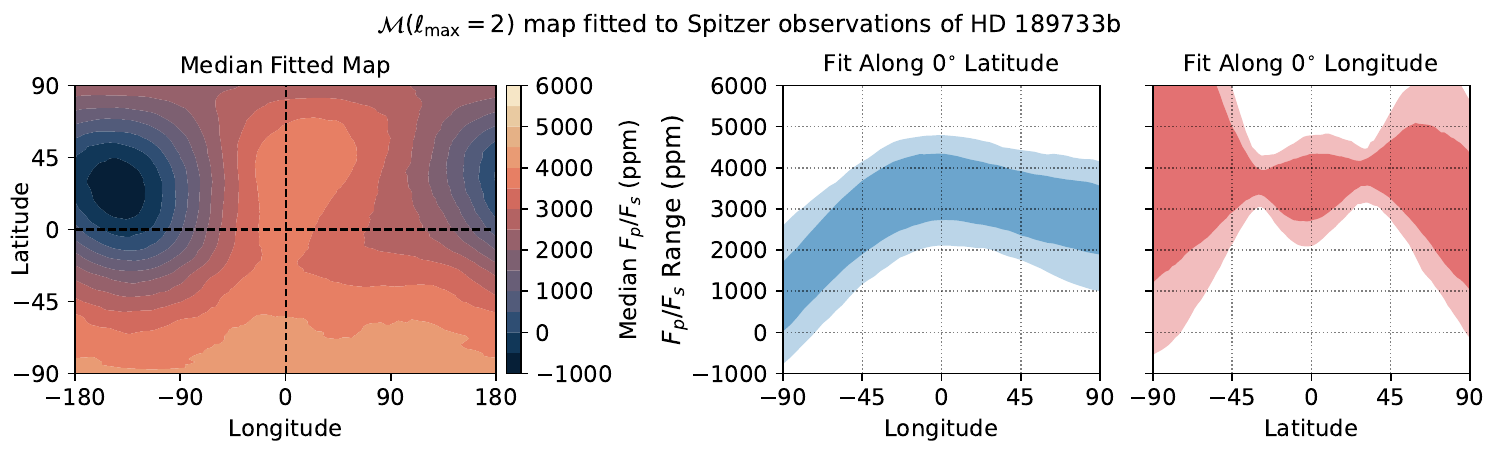}%
\caption{Supplementary Figure: an eclipse map model $\mathcal{M}(\ell_{\mathrm{max}}=2)$ fitted to the HD 189733b Spitzer observations. Table \ref{tab:obs_fit_stats} shows how this model achieves a better $\chi^{2}$ value than a Fourier series model that does not fit an eclipse map. However, Section \ref{sec:obs:identify_signals} and Figure \ref{fig:obs_cv_compare} use metrics that penalise overfitting to suggest that this model is not in fact preferred over the Fourier series model.}\label{fig:s6_hd189}
\end{figure*}


\bsp	
\label{lastpage}
\end{document}